\documentclass[prd,eqsecnum,twocolumn,amsfonts,amssymb]{revtex4}

\usepackage{graphicx}

\usepackage{bm}

\newcommand{\beq}{\begin{equation}}
\newcommand{\eeq}{\end{equation}}
\newcommand{\beqs}{\begin{eqnarray}}
\newcommand{\eeqs}{\end{eqnarray}}

\begin{document}

\title{Scheme-Independent Series for Anomalous Dimensions of
 Higher-Spin Operators at an Infrared Fixed Point in a Gauge Theory}

\author{Thomas A. Ryttov$^a$ and Robert Shrock$^b$}

\affiliation{(a) \ CP$^3$-Origins and Danish Institute for Advanced Study \\
Southern Denmark University, Campusvej 55, Odense, Denmark}

\affiliation{(b) \ C. N. Yang Institute for Theoretical Physics and
Department of Physics and Astronomy, \\
Stony Brook University, Stony Brook, New York 11794, USA }

\begin{abstract}

  We consider an asymptotically free vectorial gauge theory, with gauge group
  $G$ and $N_f$ fermions in a representation $R$ of $G$, having an infrared
  fixed point of the renormalization group. We calculate scheme-independent
  series expansions for the anomalous dimensions of higher-spin bilinear
  fermion operators at this infrared fixed point up to $O(\Delta_f^3)$, where
  $\Delta_f$ is an $N_f$-dependent expansion variable. Our general results are
  evaluated for several special cases, including the case $G={\rm SU}(N_c)$
  with $R$ equal to the fundamental and adjoint representations.

\end{abstract}

\maketitle


\section{Introduction}
\label{intro_section}

An asymptotically free gauge theory with sufficiently many massless fermions
evolves from the deep ultraviolet (UV) to an infrared fixed point (IRFP) of the
renormalization group at a zero of the beta function.  The theory at this IRFP
exhibits scale-invariance due to the vanishing of the beta function.  The
properties of the theory at this IRFP are of fundamental field-theoretic
interest. Among the basic properties are the anomalous dimensions
$\gamma^{({\cal O})}_{IR}$ of various gauge-invariant operators ${\cal O}$.

In this paper we consider an asymptotically free vectorial gauge theory of this
type, with a general gauge group $G$ and $N_f$ copies (``flavors'') of massless
Dirac fermions $\psi_i$, $i=1,...,N_f$, transforming according to a
representation $R$ of $G$ \cite{fm}.
We present scheme-independent series expansions of the anomalous dimensions of
gauge-invariant higher-spin operators that are bilinear in the fermion fields,
up to $O(\Delta_f^3)$ inclusive, at the infrared fixed point, where $\Delta_f$
is an $N_f$-dependent expansion variable defined below, in Eq. (\ref{deltaf}).
The operators that we consider are of the form (suppressing flavor indices)
$\bar\psi \gamma_{\mu_1} D_{\mu_2}...D_{\mu_j} \psi$ and $\bar\psi
\sigma_{\lambda\mu_1}D_{\mu_2}...D_{\mu_j}\psi$, where $D_\mu$ is the covariant
derivative for the gauge theory, and it is understood here and below that 
the operators are symmetrized over the Lorentz indices 
$\mu_i$, $1 \le i \le j$ and have Lorentz traces subtracted, and 
$\sigma_{\lambda\mu_1}$ is the commutator of two Dirac
matrices (defined in Eq. (\ref{sigma_munu})). 
We consider the cases $1 \le j \le 3$.

The operators $\bar\psi \gamma_{\mu_1} D_{\mu_2}...D_{\mu_j} \psi$ were
considered early on in the analysis of approximate Bjorken scaling in deep
inelastic lepton scattering and the associated development of the theory of
quantum chromodynamics (QCD). We briefly review this background
\cite{ope_gen}-\cite{gross75}. In Euclidean quantum field theory, the
short-distance operator product expansion (OPE) expresses the product of two
operators $A(x)$ and $B(y)$ as a sum of local operators ${\cal O}_i$ multiplied
by coefficient functions $c_{{\cal O}_i}$,
\beq
A(x)B(y) = \sum_i c_{{\cal O}_i}(x-y) \, {\cal O}_i((x+y)/2) \ , 
\label{sd_ope}
\eeq
in the limit where $x-y \to 0$.  Let us denote the Maxwellian (i.e.,
free-field) dimension of an operator ${\cal O}$ in mass units as $d_{\cal O}$.
Then the (free-field) dimension of the coefficient function is $d_{{c_{\cal
      O}}_i} = d_A+d_B-d_{{\cal O}_i}$, so
\beq
c_{{\cal O}_i}(x-y) \sim |x-y|^{d_{{\cal O}_i} - d_A - d_B} \ , 
\label{c_sd_ope}
\eeq
where $|x-y|$ refers to the Euclidean distance. Hence, in the short-distance
OPE, the operators with the lowest dimensions dominate, since they are
multiplied by the smallest powers of $|x-y|$. However, deep inelastic
scattering and the associated Bjorken limit probe the light cone limit,
$(x-y)^2 \to 0$ with $x-y \ne 0$ in Minkowski space, where $x^2 = x_\mu x^\mu$.
With the arguments of two illustrative Lorentz-scalar operators denoted in a
symmetric manner as $\pm x/2$, the light-cone OPE for $A(x/2)B(-x/2)$ is
\beq
A(x/2)B(-x/2) = \sum_{i,n} \bar c_{i,n}(x^2) \, x^{\mu_1} \cdots x^{\mu_n} \, 
{\cal O}_{i,n;\mu_1,...,\mu_n}(0)
\label{lc_ope}
\eeq
in the limit $x^2 \to 0$, where the coefficient functions have been written in
a form that explicitly indicates the factor $x^{\mu_1} \cdots x^{\mu_n}$ and
the operator ${\cal O}_{i,n;\mu_1,...,\mu_n}$ has spin $j=n$.  Here
(suppressing the Lorentz indices on ${\cal O}_{i,n;\mu_1,...,\mu_n}$) the
dependence of $\bar c_{i,n}$ on $x^2$ is
\beq
\bar c_{i,n}(x^2) \sim (x^2)^{(d_{{\cal O}_{i,n}} -n - d_A - d_B)/2} 
\label{cbar}
\eeq
(with logarithmic corrections in QCD due to anomalous dimensions). 
Consequently, the operators that have the strongest singularity in their
coefficient function $\bar c_{i,n}(x^2)$ as $x^2 \to 0$ and hence make the 
dominant contribution to the right-hand side of the light-cone OPE, Eq. 
(\ref{lc_ope}) are those with minimal ``twist'' $\tau$ \ \cite{gt71}, where 
$\tau$ is the dimension minus the spin $j$ of the operator, i.e., 
\beq
\tau_{{\cal O}_{i,n}} = d_{{\cal O}_{i,n}} - j_{{\cal O}_{i,n}} \ , 
\label{tau}
\eeq
with $j_{{\cal O}_{i,n}}=n$ here.  Thus, among bilinear fermion operators, in
addition to $\bar\psi\gamma_{\mu_1} \psi$ with dimension 3, spin 1, and hence
$\tau=2$, there are the operators $\bar\psi \gamma_{\mu_1} D_{\mu_2} \cdots
D_{\mu_j} \psi$, with dimension $3+(j-1)$ and spin $j$, which also have
$\tau=2$. These are the minimum-twist bilinear fermion operators that
contribute to the light-cone OPE (\ref{lc_ope}) \cite{ff}. In a similar manner,
twist-2 operators make the dominant contribution to the right-hand side of the
light-cone OPE for the product of two electromagnetic or weak currents. The
other operators that we consider, namely
$\bar\psi\sigma_{\lambda\mu_1}D_{\mu_2}...D_{\mu_j}\psi$, have been relevant
for the study of transversity distributions in QCD \cite{jj}.

Our approach here is complementary to these previous analyses of higher-spin
operators, which have focused on applications to QCD. In contrast, we study the
anomalous dimensions of these operators at an infrared fixed point in a
(deconfined) chirally symmetric non-Abelian Coulomb phase (NACP), where the
theory is scale-invariant and is inferred to be conformally invariant
\cite{scalecon}, whence the commonly used term ``conformal window''.  The goal
of our calculations is to gain information about the properties of the
conformal field theory that is defined at this IRFP.

Let us recall some further relevant background for our work. The
evolution of the running gauge coupling $g=g(\mu)$, as a function of the
momentum scale, $\mu$, is described by the renormalization-group (RG)
beta function $\beta = d\alpha/d\ln\mu$, where $\alpha(\mu) = g(\mu)^2/(4\pi)$.
From the one-loop term in the beta function \cite{b1,b1p}, it
follows that the property of asymptotic freedom restricts $N_f$ to be less than
an upper ($u$) bound, $N_u$, where \cite{nfintegral}
\beq
N_u = \frac{11C_A}{4T_f} \ .
\label{Nfb1z}
\eeq
Here, $C_A$ is the quadratic Casimir invariant for the group $G$ and $T_f$ is
the trace invariant for the representation $R$ \cite{group_invariants}.  If
$N_f$ is slightly less than $N_u$, then this theory has an infrared zero in the
(perturbatively calculated) beta function, i.e., an IR fixed point of the
renormalization group, at a value that we shall denote $\alpha_{IR}$
\cite{b2,bz}. In the two-loop beta function (with $N_f < N_u$ as required by
asymptotic freedom), this IR zero is present if $N_f$ is larger than a lower
($\ell$) value $N_\ell$, where \cite{b2}
\beq
N_\ell = \frac{17 C_A^2}{2T_f(5C_A+3C_f)} \ . 
\label{Nfb2z}
\eeq
As the scale $\mu$ decreases from large values in the UV to small values in the
IR, $\alpha(\mu)$ approaches $\alpha_{IR}$ from below as $\mu \to 0$.  Here we
consider the properties of the theory at this IRFP in the perturbative beta
function. (For a discussion of an IR zero in a nonperturbatively defined beta
function and its application to QCD, see \cite{dbt}.)

Since the anomalous dimensions of gauge-invariant operators evaluated at the
IRFP are physical, they must be independent of the scheme used for
regularization and renormalization.  In the conventional approach, one first
expresses these anomalous dimensions as series expansions in powers of $\alpha$
or equivalently $a=g^2/(16\pi^2) = \alpha/(4\pi)$, calculated to $n$-loop
order; second, one computes the IR zero of the beta function, denoted
$\alpha_{IR,n}$, to the same $n$-loop order; and third, one sets
$\alpha=\alpha_{IR,n}$ in the series expansion for the given anomalous
dimension to obtain its value at the IR zero of the beta function to this
$n$-loop order.  For the operator $\bar \psi \psi$ this conventional approach
to calculate anomalous dimensions at an IR fixed point was carried out to the
four-loop level in \cite{bvh,ps,bc} and to the five-loop level in \cite{flir}.
However, these conventional series expansions in powers of $\alpha$, calculated
to a finite order, are scheme-dependent beyond the leading terms.  This is a
well-known property of higher-order QCD calculations used to fit actual
experimental data, which, in turn, has motivated many studies to reduce scheme
dependence \cite{qcd_schemes}. These studies dealt with the UV fixed point
(UVFP) at $\alpha=0$, as is appropriate for QCD.  Studies of scheme dependence
of quantities calculated in a conventional manner at an IR fixed point at
$\alpha_{IR}$ were carried out in \cite{scc}-\cite{schl}. In particular, it was
shown that many scheme transformations that are admissible in the vicinity of
the UVFP at $\alpha=0$ in an asymptotically free theory are not admissible away
from the origin because of various pathological properties. For example, the
scheme transformation $r a = \tanh (r a')$ (depending on a parameter $r$) is an
admissible transformation in the neighborhood of $\alpha=\alpha'=0$. However,
the inverse of this transformation is $a'=(2r)^{-1}\ln[(1+ra)/(1-ra)]$, which
is singular at an IRFP with $a_{IR} \ge 1/r$, i.e., $\alpha_{IR} \ge 4\pi/r$,
so that the transformation is not admissible at this IRFP. Refs. \cite{scc}
derived and studied an explicit scheme transformation that removes terms of
loop order 3 and higher from the beta function in the local vicinity of
$\alpha=0$, as is relevant to the UVFP in QCD \cite{thooft77}, but also showed
that such a scheme transformation cannot, in general be used at an IRFP away
from the origin owing to various pathologies, one of which was illustrated
above. 

It is thus desirable to use a theoretical framework in which the series
expansions for physical quantities, such as anomalous dimensions of
gauge-invariant operators at the IRFP, are scheme-independent at any finite
order in an expansion variable. Because $\alpha_{IR} \to 0$ as $N_f$ approaches
$N_u$ from below (where $N_f$ is formally generalized here from a non-negative
integer to a non-negative real number \cite{nfintegral}), one can reexpress the
expansions for physical quantities at the IRFP as power series in the
manifestly scheme-independent quantity \cite{bz,gkgg} 
\beq
\Delta_f = N_u - N_f \ . 
\label{deltaf}
\eeq
In previous work we have calculated scheme-independent expansions for anomalous
dimensions of several types of gauge-invariant operators at an IRFP in an
asymptotically free gauge theory \cite{gtr}-\cite{cgs}.  We have compared the
resultant values for anomalous dimensions with lattice measurements where
available \cite{gsi}-\cite{dexl},\cite{lgtreviews,simons}. 

In the present paper we extend these calculations to the case of the
higher-spin operators $\bar\psi \gamma_{\mu_1} D_{\mu_2}...D_{\mu_j}\psi$ and
$\bar\psi \sigma_{\lambda\mu_1}D_{\mu_2}...D_{\mu_j}\psi$ for $1 \le j \le 3$.
In addition to general formulas, we present results for several different
special cases, including the case where $G={\rm SU}(N_c)$ and the fermions are
in the fundamental ($F$) and adjoint ($Adj$) representations. We also give
results for the limit $N_c \to \infty$ and $N_f \to \infty$ with the ratio
$N_f/N_c$ fixed and finite.  Our calculations show that these
scheme-independent expansions of the anomalous dimensions of the operators are
reasonably accurate throughout much of the non-Abelian Coulomb phase.  Our
results give further insight into the properties of a theory at an IRFP and
should be useful to compare with lattice measurements of the anomalous
dimensions of these higher-spin operators when such measurements will be
performed \cite{lgt_su3nf12}.

This paper is organized as follows. Some relevant background and methods are
discussed in Sect. \ref{methods_section}. General structural forms for the
anomalous dimensions of higher-spin bilinear fermion operators are given in
Sect. \ref{general_structure_section}. In Sect. \ref{general_kappa_section} we
present our scheme-independent calculations of the anomalous dimensions of
these higher-spin Wilson operators for a general gauge group $G$ and fermion
representation $R$.  In Sect. \ref{kappa_sun_section} we give results for the
case where $G={\rm SU}(N_c)$ and $R$ is the fundamental representation, and in
Sect. \ref{gamma_op_lnn_section} we present the special case of these results
for the limit $N_c \to \infty$ and $N_f \to \infty$ with $N_f/N_c$ fixed and
finite.  Anomalous dimension calculations for the case where $G={\rm SU}(N_c)$
and $R$ is the adjoint representation are presented in
Sect. \ref{gamma_adj_section}.  Our conclusions are given in
Sect. \ref{conclusion_section} and some auxiliary results are included in
Appendix \ref{gammat_su3_appendix}.


\section{Calculational Methods} 
\label{methods_section}

Let us consider a (gauge-invariant) operator ${\cal O}$. 
Because of the interactions, the full scaling
dimension of this operator, denoted $D_{\cal O}$, differs from its free-field 
value, $D_{\cal O,{\rm free}} \equiv d_{\cal O}$:
\beq
D_{\cal O} = D_{\cal O,{\rm free}} - \gamma_{\cal O} \ , 
\label{Dgamma} 
\eeq
where $\gamma_{\cal O}$ is the anomalous dimension of the operator
\cite{gammaconvention}.  Since $\gamma_{\cal O}$ arises from
the gauge interaction, it can be expressed as the power series
\beq
\gamma^{({\cal O})} = \sum_{\ell=1}^\infty 
c^{({\cal O})}_{\gamma,\ell} \, a^\ell \ , 
\label{gammaseries}
\eeq
where $c^{{\cal O}}_{\gamma,\ell}$ is the $\ell$-loop coefficient. 

As stated in the introduction, we shall consider the gauge-invariant operators 
${\cal O}_{\mu_1...\mu_j} = 
\bar\psi \gamma_{\mu_1}D_{\mu_2}...D_{\mu_j} \psi$ and
${\cal O}_{\lambda \mu_1...\mu_j} = 
\bar\psi \sigma_{\lambda \mu_1}D_{\mu_2}...D_{\mu_j} \psi$, where 
\beq
\sigma_{\lambda\mu_1} = \frac{i}{2}[\gamma_\lambda, \gamma_{\mu_1} ] \ . 
\label{sigma_munu}
\eeq
We focus on the operators with $1 \le j \le 3$. 
We introduce the following compact notation for these operators:
\beq
{\cal O}^{(\gamma D )}_{\mu_1 \mu_2} \equiv 
\bar \psi \gamma_{\mu_1} D_{\mu_2} \psi
\label{op_gd}
\eeq
\beq
{\cal O}^{(\gamma DD )}_{\mu_1 \mu_2 \mu_3 } \equiv 
\bar \psi \gamma_{\mu_1} D_{\mu_2} D_{\mu_3} \psi
\label{op_gdd}
\eeq
\beq
{\cal O}^{(\gamma DDD)}_{\mu_1 \mu_2 \mu_3 \mu_4 } \equiv 
\bar \psi \gamma_{\mu_1} D_{\mu_2} D_{\mu_3} D_{\mu_4}\psi
\label{op_gddd}
\eeq
\beq
{\cal O}^{(\sigma D )}_{\lambda\mu_1\mu_2} \equiv
\bar \psi \sigma_{\lambda\mu_1} D_{\mu_2} \psi
\label{op_sd}
\eeq
\beq
{\cal O}^{(\sigma DD )}_{\lambda\mu_1\mu_2\mu_3} \equiv
\bar \psi \sigma_{\lambda\mu_1} D_{\mu_2} D_{\mu_3} \psi 
\label{op_sdd}
\eeq
and
\beq
{\cal O}^{(\sigma DDD )}_{\lambda\mu_1\mu_2\mu_3} \equiv
\bar \psi \sigma_{\lambda\mu_1} D_{\mu_2} D_{\mu_3}D_{\mu_4} \psi \ . 
\label{op_sddd}
\eeq
For brevity of notation, we suppress the flavor indices on the fields $\psi$.

For a given operator ${\cal O}$, we write the scheme-independent expansion of
its anomalous dimension $\gamma^{({\cal O})}$ evaluated at the IRFP, denoted
$\gamma^{({\cal O})}_{IR}$, as
\beq
\gamma^{({\cal O})}_{IR} = \sum_{n=1}^\infty \kappa^{({\cal O})}_n \, 
\Delta_f^{\ n}  \ .
\label{gamma_delta_series}
\eeq
The truncation of right-hand side of Eq. (\ref{gamma_delta_series})
to maximal power $p$ is denoted 
\beq
\gamma^{({\cal O})}_{IR,\Delta_f^p} = \sum_{n=1}^p \kappa^{({\cal O})}_n \, 
\Delta_f^{\ n}  \ .
\label{gamma_delta_series_powerp}
\eeq
We use a further shorthand notation for the anomalous dimensions in which the
superscript in $\gamma^{({\cal O})}_{IR}$ is replaced by a symbol for the
quantity standing between $\bar\psi$ and $\psi$ in the operator ${\cal
  O}$. These shorthand symbols are as follows: $\gamma^{(\gamma D)}_{IR}$ 
for the anomalous dimension of the operator 
${\cal O}^{(\gamma D)}_{\mu_1\mu_2}= \bar\psi \gamma_{\mu_1} D_{\mu_2} \psi$ 
at the IRFP,
and so forth for the other operators.  In comparing with our previous
calculations in \cite{gtr}-\cite{dexo}, we also use the notation 
$\gamma^{(1)}_{IR}$
and $\gamma^{(\sigma)}_{IR}$ for the anomalous dimensions of $\bar\psi\psi$ and
$\bar\psi \sigma_{\lambda\mu_1}\psi$ at the IRFP. (The anomalous dimension
$\gamma^{(\sigma)}_{IR}$ was denoted $\gamma_{T,IR}$ in \cite{dex}, where 
the subscript $T$ referred to the Dirac tensor $\sigma_{\mu\nu}$.)

As discussed in \cite{gtr,dex}, the calculation of the coefficient
$\kappa^{({\cal O})}_n$ in Eq. (\ref{gamma_delta_series}) requires, as inputs,
the beta function coefficients at loop order $1 \le \ell \le n+1$ and the
anomalous dimension coefficients $c^{({\cal O})}_{\gamma,\ell}$ at loop order
$1 \le \ell \le n$. The method of calculation requires that the IR fixed point
must be exact, which is the case in the non-Abelian Coulomb phase.  As in our
earlier work \cite{gtr}-\cite{dexo}, we thus restrict our consideration to the
non-Abelian Coulomb phase (conformal window) \cite{sksb}.  For a given gauge
group $G$ and fermion representation $R$, the conformal window extends from an
upper end at $N_f=N_u$ to a lower end at a value that is commonly denoted
$N_{f,cr}$. In contrast to the exactly known value of $N_u$ (given in
Eq. (\ref{Nfb1z})), the value of $N_{f,cr}$ is not precisely known and has been
investigated extensively for several groups $G$ and fermion representations $R$
\cite{lgtreviews,simons,sksb}. For values of $N_f$ in the non-Abelian Coulomb
phase such that $\Delta_f$ is not too large, one may expect the expansion
(\ref{gamma_delta_series}) of $\gamma^{({\cal O})}_{IR}$ in a series in powers
of $\Delta_f$ to yield reasonably accurate perturbative calculations of the
anomalous dimension. In our earlier works, using our explicit calculations, we
have shown that this is, in fact, the case.

We recall some relevant properties of the theory regarding
global flavor symmetries.  Because the $N_f$ fermions are massless, the
Lagrangian is invariant under the classical global flavor ($fl$)
symmetry $G_{fl,cl} = {\rm U}(N_f)_L \otimes {\rm U}(N_f)_R$, or equivalently, 
\beqs
&& G_{fl,cl} = {\rm SU}(N_f)_L \otimes {\rm SU}(N_f)_R \otimes 
{\rm U}(1)_V \otimes {\rm U}(1)_A \cr\cr
& & 
\label{gfl_classical}
\eeqs
(where $V$ and $A$ denote vector and axial-vector). The U(1)$_V$ represents
fermion number, which is conserved by the bilinear operators that we consider.
The ${\rm U}(1)_A$ symmetry is broken by instantons, so
the actual nonanomalous global flavor symmetry is 
\beq
G_{fl} = {\rm SU}(N_f)_L \otimes {\rm SU}(N_f)_R \otimes 
{\rm U}(1)_V \ . 
\label{gfl}
\eeq
This $G_{fl}$ symmetry is respected in the non-Abelian Coulomb phase, since
there is no spontaneous chiral symmetry breaking in this phase
\cite{lgtreviews,simons}.  For our operators, the flavor matrix between $\bar
\psi$ and $\psi$ is either the identity or the operator $T_a$, a generator of
SU($N_f$), which can be viewed as acting either to the right on $\psi$ or to
the left on $\bar \psi$. These yield the same anomalous dimensions
\cite{gracey_gammatensor}. As a consequence of the unbroken global flavor
symmetry, our operators transform as representations of the global flavor group
$G_{fl}$. The invariance under the full $G_{fl}$ in the non-Abelian Coulomb
phase is different from the situation in the QCD-like phase at
smaller $N_f$, where the chiral part of $G_{fl}$ is spontaneously broken by the
QCD bilinear quark condensate to the vectorial subgroup ${\rm SU}(N_f)_V$ and
operators are classified according to whether they are singlet or nonsinglet
(adjoint) under this vectorial SU($N_f$) symmetry.  In particular, in the
consideration of flavor-singlet operators, in QCD, one must take into account
mixing with gluonic operators.  Here, in contrast, there is no mixing between
any of our bilinear fermion operators and gluonic operators, since the latter
are singlets under $G_{fl}$.

The operators ${\cal O}$ with an even number of Dirac $\gamma$ matrices,
symbolically denoted $\Gamma_e$, link left with right chiral components of
$\psi$, while the operators ${\cal O}$ with an odd number of Dirac $\gamma$
matrices, $\Gamma_o$, link left with left and right with right components:
\beq
\bar\psi \Gamma_e \psi = \bar\psi_L \Gamma_e \psi_R + 
                         \bar\psi_R \Gamma_e \psi_L
\label{operator_even}
\eeq
\beq
\bar\psi \Gamma_o \psi = \bar\psi_L \Gamma_o \psi_L + 
                         \bar\psi_R \Gamma_o \psi_R \ , 
\label{operator_odd}
\eeq
where $\bar \psi = \psi^\dagger \gamma_0$. 
In the non-Abelian Coulomb phase where the flavor symmetry is
(\ref{gfl}), one may regard the $T_b$ in the term $\bar\psi_L T_b \psi_R$ 
acting to the right as an element of ${\rm SU}(N_f)_R$ and acting to the 
left as an element of ${\rm SU}(N_f)_L$.

Given that the theory at the IR fixed point is conformally invariant
\cite{scalecon}, there is an important lower bound on the full dimension of an
operator ${\cal O}$ and hence, with our definition (\ref{Dgamma}), an upper
bound on the anomalous dimension $\gamma^{({\cal O})}$ that follows from the
conformal invariance.  To state this, we first recall that a
(finite-dimensional) representation of the Lorentz group is specified by the
set $(j_1,j_2)$, where $j_1$ and $j_2$ take on nonnegative integral or
half-integral values \cite{lorentzgroup}.  A Lorentz scalar operator thus
transforms as $(0,0)$, a Lorentz vector as $(1/2,1/2)$, an antisymmetric tensor
like the field-strength tensor $F^a_{\mu\nu}$ as $(1,0) \oplus (0,1)$, etc.
Then the requirement of unitarity in a conformally invariant theory implies the
lower bound \cite{gammabound}
\beq
D_{\cal O} \ge j_1 + j_2 + 1 \ , 
\label{dlowerbound}
\eeq
i.e., the upper bound 
\beq
\gamma_{\cal O} \le D_{\cal O, {\rm free}} - (j_1+j_2+1) \ . 
\label{gamma_upperbound}
\eeq
We have studied the constraints from the upper bound (\ref{gamma_upperbound}) 
in our previous calculations of anomalous dimensions in 
\cite{bvh,flir}, \cite{dex}-\cite{dexo}. Anticipating the results given below,
since our calculations yield negative values for the anomalous dimensions of 
higher-spin Wilson operators, they obviously satisfy these conformality upper 
bounds. 


\section{Some General Structural Properties of $\gamma^{({\cal O})}_{IR}$} 
\label{general_structure_section}

From our previous calculations \cite{gtr}-\cite{dexo} for the anomalous
dimensions of $\bar\psi \psi$ and $\bar\psi \sigma_{\mu\nu} \psi$, in
conjunction with our new results on the anomalous dimensions 
$\gamma^{({\cal O})}_{IR}$ of higher-spin twist-2 bilinear fermion
operators ${\cal O}$, we find some general structural properties of the
coefficients $\kappa^{({\cal O})}_n$ in the scheme-independent series
expansions of the anomalous dimensions $\gamma^{({\cal O})}_{IR}$.
These involve
various group invariants, including the quadratic Casimir invariants $C_A
\equiv C_2(G)$, $C_f \equiv C_2(R)$, the trace invariant $T(R)$, and the
quartic trace invariants $d_R^{abcd} d_{R'}^{abcd}/d_A$, where $d_A$ denotes
the dimension of the adjoint representation
\cite{group_invariants,rsv_invariants}. For compact notation, it is convenient
to define a factor that occurs in the denominators of these
$\kappa^{({\cal O})}_n$ coefficients, namely. 
\beq
D = 7C_A+11C_f \ .
\label{d}
\eeq
(not to be confused with covariant derivative). 
We exhibit this general form here, using 
$a^{({\cal O})}_{j,k}$ for various (constant) numerical coefficients: 
\beq
\kappa^{({\cal O})}_1 = c^{({\cal O})}_1 \frac{C_f T_f}{C_A D} \ , 
\label{kappa1_general}
\eeq
\bigskip
\beq
\kappa^{({\cal O})}_2 = \frac{C_f T_f^2(
a^{({\cal O})}_{2,1} C_A^2 + 
a^{({\cal O})}_{2,2} C_A C_f + 
a^{({\cal O})}_{2,3} C_f^2)}{C_A^2 D^3} \ , 
\label{kappa2_general}
\eeq
\begin{widetext}
and
\beqs
\kappa^{({\cal O})}_3 &=& \frac{C_f T_f}{C_A^4 D^5} \, \bigg [
a^{({\cal O})}_{3,1} C_A^5       T_f^2 + 
a^{({\cal O})}_{3,2} C_A^4 C_f   T_f^2 + 
a^{({\cal O})}_{3,3} C_A^3 C_f^2 T_f^2 +
a^{({\cal O})}_{3,4} C_A^2 C_f^3 T_f^2 +
a^{({\cal O})}_{3,5} C_A C_f^4   T_f \cr\cr
&+&
a^{({\cal O})}_{3,6} C_A T_f^2    \frac{d_A^{abcd}d_A^{abcd}}{d_A} +
a^{({\cal O})}_{3,7} C_f T_f^2    \frac{d_A^{abcd}d_A^{abcd}}{d_A} +
a^{({\cal O})}_{3,8} C_A^2 T_f    \frac{d_R^{abcd}d_A^{abcd}}{d_A} + 
a^{({\cal O})}_{3,9} C_A C_f T_f  \frac{d_R^{abcd}d_A^{abcd}}{d_A} \cr\cr
&+&
a^{({\cal O})}_{3,10} C_A^3       \frac{d_R^{abcd}d_R^{abcd}}{d_A} +
a^{({\cal O})}_{3,11} C_A^2 C_f   \frac{d_R^{abcd}d_R^{abcd}}{d_A} \bigg ] \ .
\label{kappa3_general}
\eeqs
\end{widetext}


\section{Anomalous Dimensions $\gamma^{({\cal O })}_{IR}$ of Higher-Spin 
Operators}
\label{general_kappa_section}


\subsection{General} 

In this section we present the results of our calculations of the coefficients
in the scheme-independent series expansions up to $O(\Delta_f^3)$ for the
various higher-spin operators considered here. As was noted above, the
calculation of the $O(\Delta_f^n)$ coefficient, $\kappa^{({\cal O})}_n$, for
the anomalous dimension of an operator ${\cal O}$ at the IRFP requires, as
inputs, the beta function coefficients at loop order $1 \le \ell \le n+1$ and
the anomalous dimension coefficients $c^{({\cal O})}_\ell$ at loop order $1 \le
\ell \le n$.  Hence, we use the beta function coefficients from one-loop up to
the four-loop level \cite{b1,b2}, \cite{b3,b4}, together with the anomalous
dimension coefficients calculated in the conventional series expansion in
powers of $a$ up to the three-loop level \cite{b1p}, \cite{gracey_gammatensor},
\cite{c4}--\cite{gnote}. The higher-order terms in the beta function and
anomalous dimensions that we use have been calculated in the $\overline{\rm
  MS}$ scheme \cite{msbar}, but our results are independent of this since they
are scheme-independent. (The beta function has actually been calculated up to
five-loop order \cite{b5su3,b5}, but these results will not be needed here.)


\subsection{ $\gamma^{(\gamma D)}_{IR}$ }

For the anomalous dimension $\gamma^{(\gamma D)}_{IR}$ of the operator 
$\bar\psi \gamma_{\mu_1} D_{\mu_2} \psi$ at the IRFP, we calculate 
\beq
\kappa^{(\gamma D)}_1 = -\frac{2^6 C_f T_f}{3^2 C_A D}  \ , 
\label{kappa1_gd}
\eeq
\beqs
\kappa^{(\gamma D)}_2 &=& \frac{2^5 C_f T_f^2 \Big ( 
693C_A^2-3104C_AC_f-1540C_f^2 \Big ) }{3^5 C_A^2 D^3} \ , \cr\cr
&&
\label{kappa2_gd}
\eeqs
and 
\begin{widetext} 
\beqs
&& \kappa^{(\gamma D)}_3 = -\frac{2^5 C_f T_f }{3^8 C_A^4 D^5} \, \bigg [ 
C_A^5T_f^2(-202419+1016064\zeta_3) +
C_A^4 C_f T_f^2(2764440+145152\zeta_3) \cr\cr
&+& 
C_A^3 C_f^2 T_f^2(8940028-5038848\zeta_3)+
C_A^2 C_f^3 T_f^2(-7341488-1140480\zeta_3)+
C_A C_f^4 T_f^2(3841024+5018112\zeta_3) \cr\cr
&+&
C_A T_f^2 \frac{d_A^{abcd}d_A^{abcd}}{d_A} (-161280+4257792\zeta_3)+
C_f T_f^2 \frac{d_A^{abcd}d_A^{abcd}}{d_A} (-253440+6690816\zeta_3) \cr\cr
&+&
C_A^2T_f \frac{d_R^{abcd}d_A^{abcd}}{d_A} (2838528-27675648\zeta_3) + 
C_AC_fT_f\frac{d_R^{abcd}d_A^{abcd}}{d_A} (4460544-43490304\zeta_3) \cr\cr
&+&
C_A^3 \frac{d_R^{abcd}d_R^{abcd}}{d_A} (-10733184+23417856\zeta_3) +
C_A^2C_f \frac{d_R^{abcd}d_R^{abcd}}{d_A} (-16866432+36799488\zeta_3)
 \bigg ] \ . \cr\cr
&&
\label{kappa3_gd}
\eeqs
\end{widetext}
In these expressions and the following ones, we have indicated the
factorizations of the numbers in the denominators, since they are rather
simple. In general, the numbers in the numerators do not have such simple
factorizations.  

With these coefficients, the anomalous dimension $\gamma^{(\gamma D)}_{IR}$
calculated to order $O(\Delta_f^p)$, denoted $\gamma^{(\gamma
  D)}_{IR,F,\Delta_f^p}$, is given by Eq. (\ref{gamma_delta_series_powerp})
with ${\cal O} = \bar\psi \gamma_{\mu_1} D_{\mu_2} \psi$. Our results here
yield $\gamma^{(\gamma D)}_{IR,F,\Delta_f^p}$ with $p=1, \ 2, \ 3$.  Analogous
statements apply to the anomalous dimensions of the other operators for which
we have performed calculations, and we proceed to present the coefficients for
these next.


\subsection{ $\gamma^{(\gamma DD)}_{IR}$ }

For the anomalous dimension $\gamma^{(\gamma DD)}_{IR}$ of the operator 
$\bar\psi \gamma_{\mu_1} D_{\mu_2}D_{\mu_3} \psi$ at the IRFP, we calculate 
\beq
\kappa^{(\gamma DD)}_1 = -\frac{100 C_f T_f}{3^2 C_A D}  \ , 
\label{kappa1_gdd}
\eeq
\beqs
\kappa^{(\gamma DD)}_2 &=& \frac{10  C_f T_f^2 \Big ( 
5103C_A^2-14017C_AC_f-9383C_f^2 \Big ) }{3^5 C_A^2 D^3} \ , \cr\cr
&&
\label{kappa2_gdd}
\eeqs
and
\begin{widetext} 
\beqs
&& \kappa^{(\gamma DD)}_3 = -\frac{10C_f T_f }{3^8 C_A^4 D^5} \, \bigg [ 
C_A^5T_f^2(1538649+2794176\zeta_3) +
C_A^4 C_f T_f^2(14860881+399168\zeta_3) \cr\cr
&+& 
C_A^3 C_f^2 T_f^2(40821518-13856832\zeta_3)+
C_A^2 C_f^3 T_f^2(-35403412-3136320\zeta_3)+
C_A C_f^4 T_f^2(19308575+13799808\zeta_3) \cr\cr
&+&
C_A T_f^2 \frac{d_A^{abcd}d_A^{abcd}}{d_A} (-806400+21288960\zeta_3)+
C_f T_f^2 \frac{d_A^{abcd}d_A^{abcd}}{d_A} (-1267200+33454080\zeta_3) \cr\cr
&+&
C_A^2T_f \frac{d_R^{abcd}d_A^{abcd}}{d_A} (14192640-138378240\zeta_3) + 
C_AC_fT_f\frac{d_R^{abcd}d_A^{abcd}}{d_A} (22302720-217451520\zeta_3) \cr\cr
&+&
C_A^3 \frac{d_R^{abcd}d_R^{abcd}}{d_A} ( -53665920+117089280\zeta_3) +
C_A^2C_f \frac{d_R^{abcd}d_R^{abcd}}{d_A} (-84332160+183997440\zeta_3)
 \bigg ] \ . \cr\cr
&&
\label{kappa3_gdd}
\eeqs
\end{widetext}
%


\subsection{ $\gamma^{(\gamma DDD)}_{IR}$ }

Proceeding to the anomalous dimension $\gamma^{(\gamma DDD)}_{IR}$ of the
operator 
$\bar\psi \gamma_{\mu_1} D_{\mu_2} D_{\mu_3} D_{\mu^4} \psi$ at the IRFP, we 
find 
\beq
\kappa^{(\gamma DDD)}_1 = -\frac{628 C_f T_f }{3^2 \cdot 5 C_A D}  \ , 
\label{kappa1_gddd}
\eeq
\beqs
\kappa^{(\gamma DDD)}_2 &=& \frac{2 C_f T_f^2 \Big ( 
4550175C_A^2-10373329C_AC_f-7719767C_f^2 \Big ) }{3^5 \cdot 5^3 C_A^2 D^3} \ , 
\cr\cr
&&
\label{kappa2_gddd}
\eeqs
and
\begin{widetext}
\beqs
&& \kappa^{(\gamma DDD)}_3 = \frac{2 C_f T_f}{3^8 \cdot 5^5 C_A^4 D^5} \, 
\bigg [ 
C_A^5T_f^2(-67181774625-45691128000 \zeta_3) +
C_A^4C_fT_f^2(-318706112025-6527304000\zeta_3) \cr\cr
&+& 
C_A^3C_f^2T_f^2(-720947009518+226590696000\zeta_3)+
C_A^2C_f^3T_f^2(709569531572+51285960000\zeta_3) \cr\cr
&+& 
C_AC_f^4T_f^2(-433168554247-225658224000\zeta_3) \cr\cr
&+&
C_AT_f^2 \frac{d_A^{abcd}d_A^{abcd}}{d_A} (15825600000-417795840000\zeta_3)+
C_fT_f^2 \frac{d_A^{abcd}d_A^{abcd}}{d_A} (24868800000-6565363200000\zeta_3) 
\cr\cr
&+& 
C_A^2T_f\frac{d_R^{abcd}d_A^{abcd}}{d_A}(-278530560000+2715672960000\zeta_3) 
+ 
C_AC_fT_f\frac{d_R^{abcd}d_A^{abcd}}{d_A}(-437690880000+4267486080000\zeta_3) 
\cr\cr
&+&
C_A^3\frac{d_R^{abcd}d_R^{abcd}}{d_A}(1053193680000-2297877120000\zeta_3) +
C_A^2C_f\frac{d_R^{abcd}d_R^{abcd}}{d_A}
(1655018640000-3610949760000\zeta_3) \bigg ] \ . \cr\cr
&&
\label{kappa3_gddd}
\eeqs
\end{widetext} 
%


\subsection{ $\gamma^{(\sigma D)}_{IR}$ }

For the anomalous dimension $\gamma^{(\sigma D)}_{IR}$ of the operator 
$\bar\psi \sigma_{\lambda\mu_1}D_{\mu_2} \psi$ at the IRFP, we calculate 
\beq
\kappa^{(\sigma D)}_1 = -\frac{8 C_f T_f }{C_A D}  \ , 
\label{kappa1_sd}
\eeq
\beqs
\kappa^{(\sigma D)}_2 &=& \frac{4 C_f T_f^2 \Big ( 
77 C_A^2 -348 C_AC_f - 176 C_f^2 \Big ) }{3 C_A^2 D^3} \ , \cr\cr
&&
\label{kappa2_sd}
\eeqs
\begin{widetext}
\beqs
&& \kappa^{(\sigma D)}_3 = \frac{4 C_f T_f}{3^4 C_A^4 D^5} \, 
\bigg [ 13083C_A^5T_f^2 -240492 C_A^4C_fT_f^2 
-819408C_A^3C_f^2T_f^2+738144C_A^2C_f^3T_f^2 -662112 C_AC_f^4T_f^2\cr\cr
&+&
C_AT_f^2 \frac{d_A^{abcd}d_A^{abcd}}{d_A} (17920-473088\zeta_3)+
C_fT_f^2 \frac{d_A^{abcd}d_A^{abcd}}{d_A} (28160-743424\zeta_3) 
\cr\cr
&+& 
C_A^2T_f \frac{d_R^{abcd}d_A^{abcd}}{d_A}(-315392 + 3075072 \zeta_3) + 
C_AC_fT_f\frac{d_R^{abcd}d_A^{abcd}}{d_A}(-495616 + 4832256 \zeta_3) 
\cr\cr
&+&
C_A^3   \frac{d_R^{abcd}d_R^{abcd}}{d_A}(1192576 -2601984 \zeta_3) +
C_A^2C_f\frac{d_R^{abcd}d_R^{abcd}}{d_A}(1874048 -4088832 \zeta_3) 
\bigg ] \ . \cr\cr
&&
\label{kappa3_sd}
\eeqs
\end{widetext} 
%


\subsection{ $\gamma^{(\sigma DD)}_{IR}$ }

For the anomalous dimension $\gamma^{(\sigma DD)}_{IR}$ we calculate 
\beq
\kappa^{(\sigma DD)}_1 = -\frac{104 C_f T_f }{3^2 C_A D}  \ , 
\label{kappa1_sdd}
\eeq
\beqs
\kappa^{(\sigma DD)}_2 &=& \frac{4 C_f T_f^2 \Big ( 
12537C_A^2 -36292C_AC_f -22352C_f^2 \Big ) }{3^5 C_A^2 D^3} \ , \cr\cr
&&
\label{kappa2_sdd}
\eeqs
\begin{widetext}
and
\beqs
&& \kappa^{(\sigma DD)}_3 = -\frac{2^2 C_f T_f}{3^8 C_A^4 D^5} \, 
\bigg [ 
C_A^5T_f^2(2935737+4064256\zeta_3) +
C_A^4C_fT_f^2(39906468+580608\zeta_3) \cr\cr
&+& 
C_A^3C_f^2T_f^2(107242456-20155392\zeta_3)+
C_A^2C_f^3T_f^2(-102128048-4561920\zeta_3) \cr\cr &+& 
C_AC_f^4T_f^2(43045024+20072448\zeta_3) \cr\cr
&+&
3C_AT_f^2 \frac{d_A^{abcd}d_A^{abcd}}{d_A} (-698880+18450432\zeta_3)+
3C_fT_f^2 \frac{d_A^{abcd}d_A^{abcd}}{d_A} (-1098240+28993536\zeta_3) 
\cr\cr
&+& 
3C_A^2T_f\frac{d_R^{abcd}d_A^{abcd}}{d_A}(12300288-119927808\zeta_3) + 
3C_AC_fT_f\frac{d_R^{abcd}d_A^{abcd}}{d_A}(19329024-188457984\zeta_3) 
\cr\cr
&+&
3C_A^3\frac{d_R^{abcd}d_R^{abcd}}{d_A}(-46510464+101477376\zeta_3) +
3C_A^2C_f\frac{d_R^{abcd}d_R^{abcd}}{d_A}
(-73087872+159464448\zeta_3) \bigg ] \ . \cr\cr
&&
\label{kappa3_sdd}
\eeqs
\end{widetext} 
%

\subsection{ $\gamma^{(\sigma DDD)}_{IR}$ }

Finally, for the anomalous dimension $\gamma^{\sigma DDD)}_{IR}$ we obtain 
\beq
\kappa^{(\sigma DDD)}_1 = -\frac{2^7 C_f T_f }{3^2 C_A D} \ , 
\label{kappa1_sddd}
\eeq
\beqs
\kappa^{(\sigma DDD)}_2 &=& \frac{2^3 C_f T_f^2 \Big ( 
9219C_A^2 -21185C_A C_f -15664C_f^2 \Big ) }{3^5 C_A^2 D^3} \ , \cr\cr
&&
\label{kappa2_sddd}
\eeqs
and
\begin{widetext}
\beqs
&& \kappa^{(\sigma DDD)}_3 = -\frac{2^3C_f T_f}{3^8 C_A^4 D^5} \, 
\bigg [ 
C_A^5T_f^2(5213502+2667168\zeta_3) +
C_A^4C_fT_f^2(25185069+381024\zeta_3) \cr\cr
&+& 
C_A^3C_f^2T_f^2(58268711-13226976\zeta_3)+
C_A^2C_f^3T_f^2(-56962840-2993760\zeta_3) \cr\cr
&+& 
C_AC_f^4T_f^2(36476660+13172544\zeta_3) \cr\cr
&+&
C_AT_f^2 \frac{d_A^{abcd}d_A^{abcd}}{d_A}(-1290240+34062336\zeta_3)+
C_fT_f^2 \frac{d_A^{abcd}d_A^{abcd}}{d_A}(-2027520+53526528\zeta_3) 
\cr\cr
&+& 
C_A^2T_f\frac{d_R^{abcd}d_A^{abcd}}{d_A}(22708224-221405184\zeta_3) + 
C_AC_fT_f\frac{d_R^{abcd}d_A^{abcd}}{d_A}(35684352-347922432\zeta_3) 
\cr\cr
&+&
C_A^3\frac{d_R^{abcd}d_R^{abcd}}{d_A}(-85865472+187342848\zeta_3) +
C_A^2C_f\frac{d_R^{abcd}d_R^{abcd}}{d_A}(-134931456+294395904\zeta_3) 
\bigg ] \ . \cr\cr
&&
\label{kappa3_sddd}
\eeqs
\end{widetext} 
%


\section{Evaluation of $\kappa^{({\cal O})}_n$ for $G={\rm SU}(N_c)$ and $R=F$}
\label{kappa_sun_section}

In this section we evaluate our general results for these anomalous dimensions
$\gamma^{({\cal O})}_{IR}$ in the important special case where 
the gauge group is $G={\rm SU}(N_c)$ and the $N_f$ fermions are in the
fundamental representation of this group, $R=F$.


\subsection{ $\gamma^{(\gamma D)}_{IR,{\rm SU}(N_c),F}$ }

Substituting $G={\rm SU}(N_c)$ and $R=F$ in our general results 
(\ref{kappa1_gd})-(\ref{kappa3_gd}), we obtain 
\beq
\kappa^{(\gamma D)}_{1,{\rm SU}(N_c),F} = -\frac{2^5(N_c^2-1)}
{3^2N_c(25N_c^2-11)}  \ , 
\label{kappa1_gd_fund}
\eeq
\beq
\kappa^{(\gamma D)}_{2,{\rm SU}(N_c),F} = 
-\frac{2^5(N_c^2-1)(1244N_c^4-2322N_c^2+385)}
{3^5 N_c^2(25N_c^2-11)^3} \ , 
\label{kappa2_gd_fund}
\eeq
and
\begin{widetext}
\beqs
\kappa^{(\gamma D)}_{3,{\rm SU}(N_c),F} &=& 
-\frac{2^6(N_c^2-1)}{3^8 N_c^3 (25N_c^2-11)^5} \, 
\bigg [ 2137786N_c^8+(1831104-9784800\zeta_3)N_c^6 \cr\cr
&+&(-15928259+36575712\zeta_3)N_c^4 
+(6282342-14911776\zeta_3)N_c^2 \cr\cr
&+&240064+313632\zeta_3 \bigg ] \ . 
\label{kappa3_gd_fund}
\eeqs
\end{widetext}
Then, for this case $G={\rm SU}(3)$, $R=F$, the anomalous dimension
$\gamma^{(\gamma D)}_{IR}$ calculated to order $O(\Delta_f^p)$, denoted
$\gamma^{(\gamma D)}_{IR,F,\Delta_f^p}$, is given by
Eq. (\ref{gamma_delta_series_powerp}) with ${\cal O} = \bar\psi \gamma_{\mu_1}
D_{\mu_2} \psi$.  


\subsection{ $\gamma^{(\gamma DD)}_{IR,{\rm SU}(N_c),F}$ }

Substituting $G={\rm SU}(N_c)$ and $R=F$ in our general results 
(\ref{kappa1_gdd})-(\ref{kappa3_gdd}), we obtain 
\beq
\kappa^{(\gamma DD)}_{1,{\rm SU}(N_c),F} = -\frac{50(N_c^2-1)}
{3^2N_c(25N_c^2-11)}  \ , 
\label{kappa1_gdd_fund}
\eeq
\beq
\kappa^{(\gamma DD)}_{2,{\rm SU}(N_c),F} = 
-\frac{5(N_c^2-1)(17005N_c^4-46800N_c^2+9383)}
{2 \cdot 3^5 N_c^2(25N_c^2-11)^3} \ , 
\label{kappa2_gdd_fund}
\eeq
and
\begin{widetext}
\beqs
\kappa^{(\gamma DD)}_{3,{\rm SU}(N_c),F} &=& 
-\frac{5(N_c^2-1)}{2^2 \cdot 3^8N_c^3(25N_c^2-11)^5} \, 
\bigg [ 207341255N_c^8+(160969860-841104000\zeta_3)N_c^6 \cr\cr
&+&(-1281330310+2919058560\zeta_3)N_c^4 
+(499565484-1152911232\zeta_3)N_c^2 \cr\cr
&+&19308575+13799808\zeta_3 \bigg ] \ . 
\label{kappa3_gdd_fund}
\eeqs
\end{widetext}
%


\subsection{ $\gamma^{(\gamma DDD)}_{IR,{\rm SU}(N_c),F}$ }

In a similar manner, from our general formulas
(\ref{kappa1_gddd})-(\ref{kappa3_gddd}), we find 
\beq
\kappa^{(\gamma DDD)}_{1,{\rm SU}(N_c),F} = -\frac{314(N_c^2-1)}
{3^2 \cdot 5 N_c(25N_c^2-11)}  \ , 
\label{kappa1_gddd_fund}
\eeq
\begin{widetext}
\beq
\kappa^{(\gamma DDD)}_{2,{\rm SU}(N_c),F} = 
-\frac{(N_c^2-1)(10265725N_c^4-36186192N_c^2+7719767)}
{2 \cdot 3^5 \cdot 5^3 N_c^2(25N_c^2-11)^3} \ , 
\label{kappa2_gddd_fund}
\eeq
and
\beqs
\kappa^{(\gamma DDD)}_{3,{\rm SU}(N_c),F} &=& 
-\frac{(N_c^2-1)}{2^2 \cdot 3^8 \cdot 5^5 N_c^3(25N_c^2-11)^5} \, 
\bigg [ 4581316819375N_c^8+ (3455659520100-16739946000000\zeta_3)N_c^6 \cr\cr
&+&
(-25230047265878+57258530640000\zeta_3)N_c^4 +
(9616576686156-22465759536000\zeta_3)N_c^2 \cr\cr
&+&
433168554247+225658224000\zeta_3) \bigg ] \ . 
\label{kappa3_gddd_fund}
\eeqs
\end{widetext}
%


\subsection{ $\gamma^{(\sigma D)}_{IR,{\rm SU}(N_c),F}$ }

From our general results (\ref{kappa1_sd})-(\ref{kappa3_sd}), we obtain 
\beq
\kappa^{(\sigma D)}_{1,{\rm SU}(N_c),F} = -\frac{4(N_c^2-1)}
{N_c(25N_c^2-11)}  \ , 
\label{kappa1_sd_fund}
\eeq
\beqs
\kappa^{(\sigma D)}_{2,{\rm SU}(N_c),F} &=& 
-\frac{4(N_c^2-1)(141 N_c^4 -262N_c^2 + 44 )}
{3 N_c^2(25N_c^2-11)^3} \ , \cr\cr
&&
\label{kappa2_sd_fund}
\eeqs
and
\begin{widetext}
\beqs
\kappa^{(\sigma D)}_{3,{\rm SU}(N_c),F} &=& 
-\frac{2^3 (N_c^2-1)}{3^3 N_c^3(25N_c^2-11)^5} \, 
\bigg [ 64843N_c^8+(78610-422400\zeta_3)N_c^6 \cr\cr
&+& (-565316+1347456\zeta_3)N_c^4 
 +(209836-511104\zeta_3)N_c^2 +13794 \bigg ] \ . 
\label{kappa3_sd_fund}
\eeqs
\end{widetext} 
%


\subsection{ $\gamma^{(\sigma DD)}_{IR,{\rm SU}(N_c),F}$ }

From our general results (\ref{kappa1_sdd})-(\ref{kappa3_sdd}), we obtain 
\beq
\kappa^{(\sigma DD)}_{1,{\rm SU}(N_c),F} = -\frac{52(N_c^2-1)}
{3^2 N_c(25N_c^2-11)}  \ , 
\label{kappa1_sdd_fund}
\eeq
\begin{widetext}
\beqs
\kappa^{(\sigma DD)}_{2,{\rm SU}(N_c),F} &=& 
-\frac{4(N_c^2-1)(11197 N_c^4 -29322N_c^2 + 5588 )}
{3^5 N_c^2(25N_c^2-11)^3} \ , \cr\cr
&&
\label{kappa2_sdd_fund}
\eeqs
and
\beqs
\kappa^{(\sigma DD)}_{3,{\rm SU}(N_c),F} &=& 
-\frac{2^3 (N_c^2-1)}{3^8 N_c^3(25N_c^2-11)^5} \, 
\bigg [ 31831693N_c^8+(30539268-141782400\zeta_3)N_c^6 \cr\cr
&+& (-214403216+473734656\zeta_3)N_c^4 
+(84228606-183845376\zeta_3)N_c^2 \cr\cr
&+&2690314+1254528\zeta_3 \bigg ] \ . 
\label{kappa3_sdd_fund}
\eeqs
\end{widetext}
%


\subsection{ $\gamma^{(\sigma DDD)}_{IR,{\rm SU}(N_c),F}$ }

For this case we have 
\beq
\kappa^{(\sigma DDD)}_{1,{\rm SU}(N_c),F} = -\frac{2^6(N_c^2-1)}
{3^2 N_c(25N_c^2-11)}  \ , 
\label{kappa1_sigma_ddd_fund}
\eeq
\begin{widetext}
\beqs
\kappa^{(\sigma DDD)}_{2,{\rm SU}(N_c),F} &=& 
-\frac{2^2(N_c^2-1)(10579 N_c^4 -36849N_c^2 +7832 )}
{3^5 N_c^2(25N_c^2-11)^3} \ , \cr\cr
&&
\label{kappa2_sigma_ddd_fund}
\eeqs
and
\beqs
\kappa^{(\sigma DDD)}_{3,{\rm SU}(N_c),F} &=& 
-\frac{2^2(N_c^2-1)}{3^8 N_c^3(25N_c^2-11)^5} \, 
\bigg [
(90949802N_c^8+
(70557192-347943600\zeta_3)N_c^6 \cr\cr
&+&
(-511679503+1166243184\zeta_3)N_c^4 +
(194401944-453269520\zeta_3)N_c^2 \cr\cr
&+&9119165+3293136\zeta_3 \bigg ] \ . 
\label{kappa3_sigma_ddd_fund}
\eeqs
\end{widetext}
Below, where the meaning is clear, we will often omit the SU(3) in the
subscript. 

We remark on the signs of these coefficients. 
It is evident from Eqs. (\ref{kappa1_gd}), (\ref{kappa1_gdd}), 
(\ref{kappa1_gddd}), (\ref{kappa1_sd}), (\ref{kappa1_sdd}), and 
(\ref{kappa1_sddd}) that $\kappa^{(\gamma D)}_1$, 
$\kappa^{(\gamma DD)}_1$, $\kappa^{(\gamma DDD)}_1$, 
$\kappa^{(\sigma D)}_1$, $\kappa^{(\sigma DD)}_1$, and
$\kappa^{(\sigma DDD)}_1$ are all negative 
for any $G$ and $R$.  We find that the $O(\Delta_f^2)$ and $O(\Delta_f^3)$
coefficients, $\kappa^{({\cal O})}_2$ and $\kappa^{({\cal O})}_3$, 
for these operators are also negative for the theory with 
$G={\rm SU}(N_c)$ and fermions in the fundamental representation, $R=F$, 
in the full range $N_c \ge 2$ of relevance here. 
In Table \ref{gamma_gddd_values} we list the signs of these coefficients 
$\kappa^{({\cal O})}_n$ for the operators in this theory. For comparison, we
also include the signs of $\kappa^{(1)}_n$ for $\bar\psi\psi$ and 
$\kappa^{(\sigma)}_n$ for $\bar\psi\sigma_{\mu\nu}\psi$ that we obtained in our
earlier calculations (which hold for all $N_c$).  

It is interesting to note that for all of the higher-spin operators ${\cal O}$
that we consider, the anomalous dimensions $\gamma^{({\cal O})}_{IR}$ that we
calculate are negative (with our sign convention in (\ref{Dgamma})
\cite{gammaconvention})). They thus have the same sign as the sign of the
anomalous dimension of the operator $\bar\psi\sigma_{\mu\nu}\psi$ and are 
opposite in sign relative to the anomalous dimensions that we calculated for
$\bar\psi\psi$ in our previous work \cite{bvh}, \cite{gtr}-\cite{dexo}.

In Tables \ref{gamma_gd_values}-\ref{gamma_sddd_values} we list values of the
anomalous dimensions $\gamma^{(\gamma D)}_{IR}$, $\gamma^{(\gamma DD)}_{IR}$,
$\gamma^{(\gamma DDD)}_{IR}$, $\gamma^{(\sigma)}_{IR}$, $\gamma^{(\sigma
  D)}_{IR}$, $\gamma^{(\sigma DD)}_{IR}$, and $\gamma^{(\sigma DDD)}_{IR}$ for
the theory with $G={\rm SU}(3)$ and fermions in the fundamental representation,
$R=F$, calculated to $O(\Delta_f^p)$, denoted $\gamma^{(\gamma
  D)}_{IR,F,\Delta_f^p}$, etc., with $p=1, \ 2, \ 3$, as functions of $N_f$ for
a relevant range of $N_f$ values extending downward from the upper end of the
conformal regime at $N_f=N_u$ (i.e., $\Delta_f=0$) within this conformal window
\cite{lowerend}.  The numbers in Table \ref{gamma_s_values}) are evaluations of
our analytic results given in \cite{dex} and are included for comparison.

\begin{figure}
  \begin{center}
    \includegraphics[height=6cm]{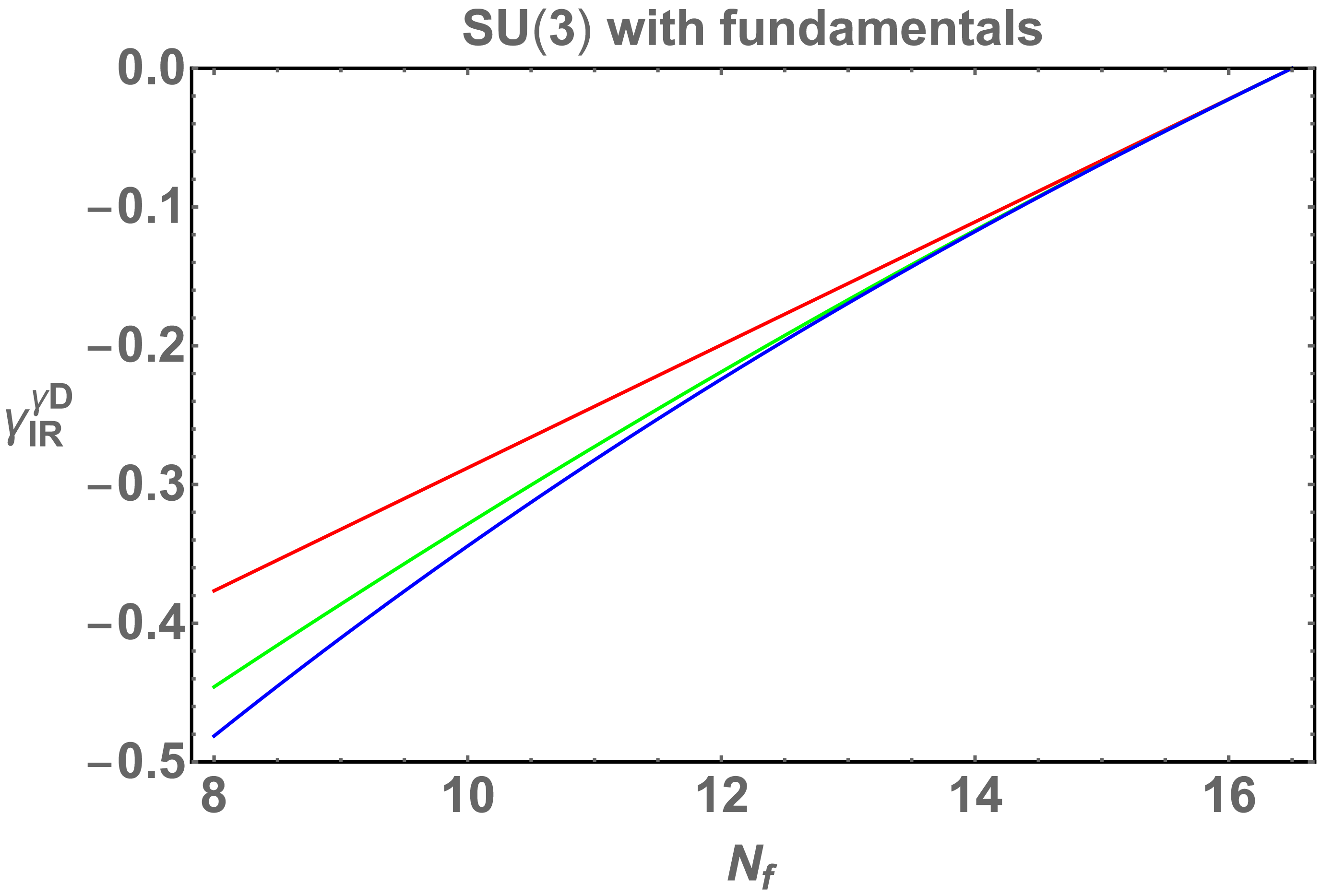}
  \end{center}
\caption{Plot of the anomalous dimension $\gamma^{(\gamma D)}_{IR,F}$ of the 
operator $\bar\psi \gamma_{\mu_1} D_{\mu_2} \psi$ at the IRFP for the theory
with $G={\rm SU}(3)$, and $N_f$ fermions in the fundamental representation,
calculated to order $O(\Delta_f^p)$, where $p=1, \ 2, \ 3$. Denoting the 
anomalous dimension calculated to order $O(\Delta_f^p)$ as 
$\gamma^{(\gamma D)}_{IR,F,\Delta_f^p}$, the curves, from top to bottom 
(with colors online), refer to 
$\gamma^{(\gamma D)}_{IR,F,\Delta_f}$ (red), 
$\gamma^{(\gamma D)}_{IR,F,\Delta_f^2}$ (green), and 
$\gamma^{(\gamma D)}_{IR,F,\Delta_f^3}$ (blue).}
\label{gd_fund_plot}
\end{figure}

\begin{figure}
  \begin{center}
    \includegraphics[height=6cm]{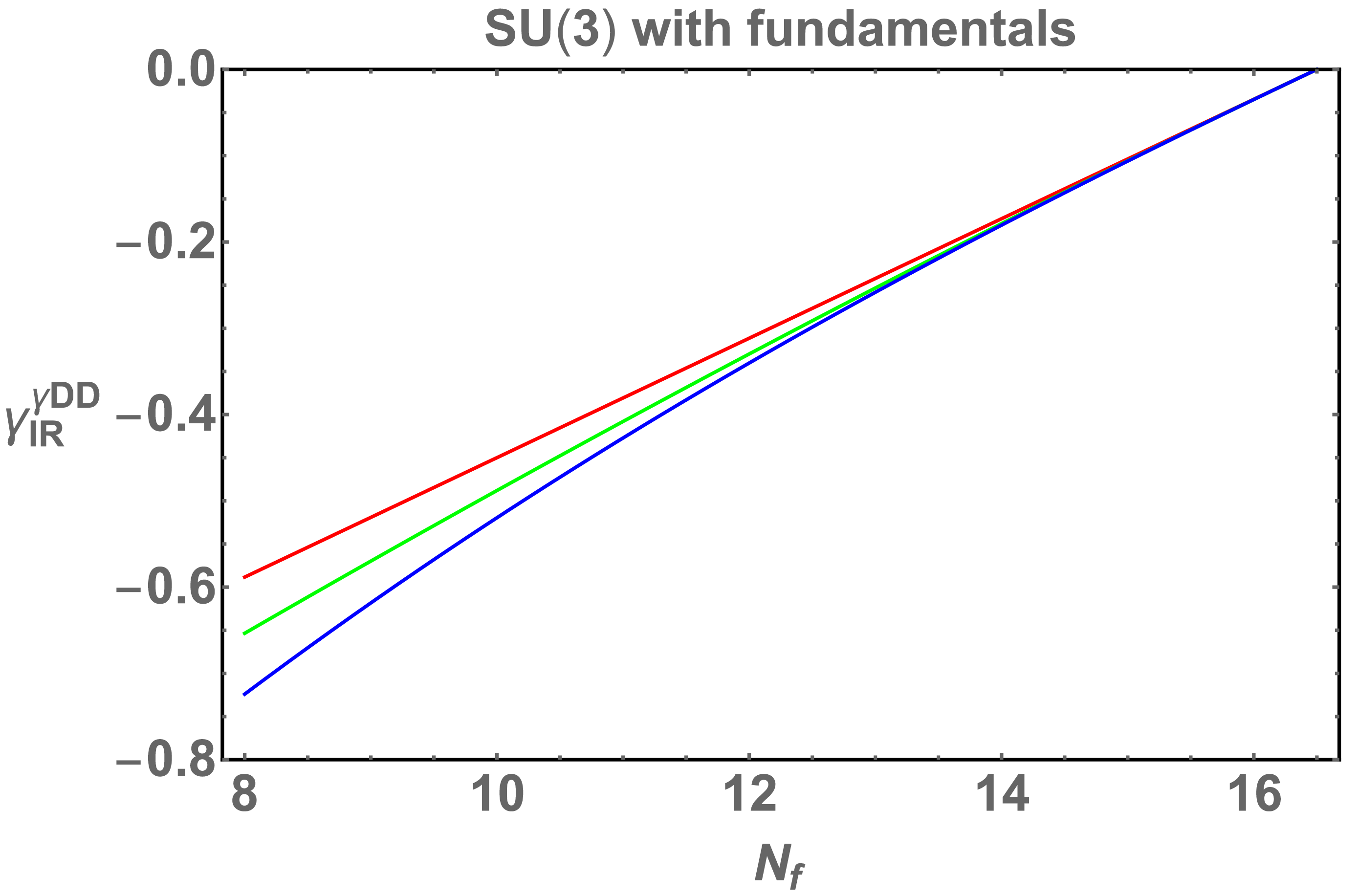}
  \end{center}
\caption{Plot of the anomalous dimension $\gamma^{(\gamma DD)}_{IR,F}$ of the 
operator $\bar\psi \gamma_{\mu_1} D_{\mu_2} D_{\mu_3} \psi$ at the IRFP for 
$G={\rm SU}(3)$, and $N_f$ fermions in the fundamental representation,
calculated to order $O(\Delta_f^p)$, where $p=1, \ 2, \ 3$. Denoting the 
anomalous dimension calculated to order $O(\Delta_f^p)$ as 
$\gamma^{(\gamma DD)}_{IR,F,\Delta_f^p}$, the curves, from top to bottom 
(with colors online), refer to 
$\gamma^{(\gamma DD)}_{IR,F,\Delta_f}$ (red), 
$\gamma^{(\gamma DD)}_{IR,F,\Delta_f^2}$ (green), and 
$\gamma^{(\gamma DD)}_{IR,F,\Delta_f^3}$ (blue).} 
\label{gdd_fund_plot}
\end{figure}

\begin{figure}
  \begin{center}
    \includegraphics[height=6cm]{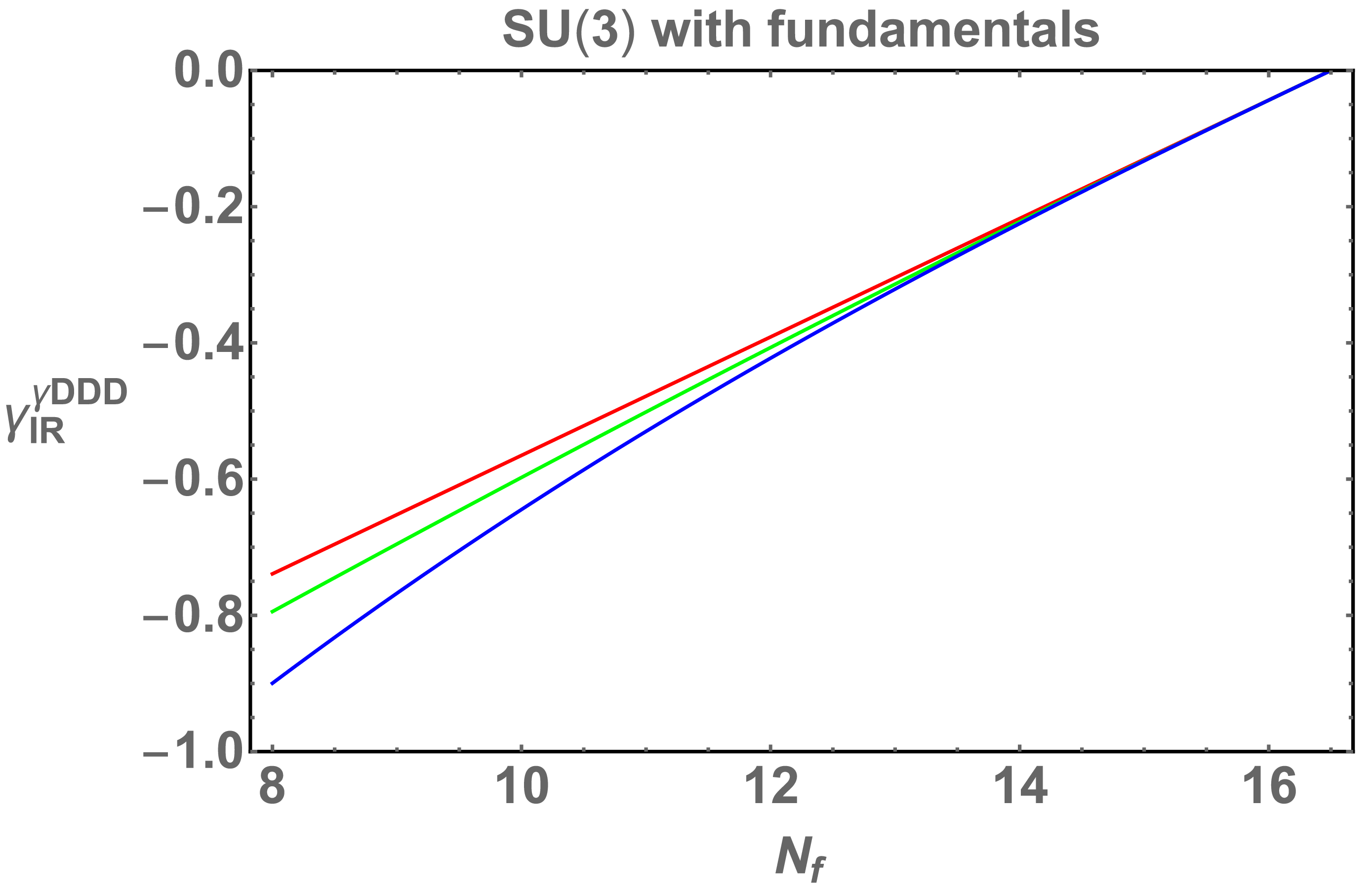}
  \end{center}
\caption{Plot of the anomalous dimension $\gamma^{(\gamma DDD)}_{IR,F}$ of the 
operator $\bar\psi \gamma_{\mu_1} D_{\mu_2} D_{\mu_3} D_{\mu_4} \psi$ at the 
IRFP for $G={\rm SU}(3)$, and $N_f$ fermions in the fundamental representation,
calculated to order $O(\Delta_f^p)$, where $p=1, \ 2, \ 3$. Denoting the 
calculation to order $O(\Delta_f^p)$ as 
$\gamma^{(\gamma DDD)}_{IR,F,\Delta_f^p}$, from top to bottom 
(with colors online), the colors refer to 
$\gamma^{(\gamma DDD)}_{IR,F,\Delta_f}$ (red), 
$\gamma^{(\gamma DDD)}_{IR,F,\Delta_f^2}$ (green), and 
$\gamma^{(\gamma DDD)}_{IR,F,\Delta_f^3}$ (blue).} 
\label{gddd_fund_plot}
\end{figure}

\begin{figure}
  \begin{center}
    \includegraphics[height=6cm]{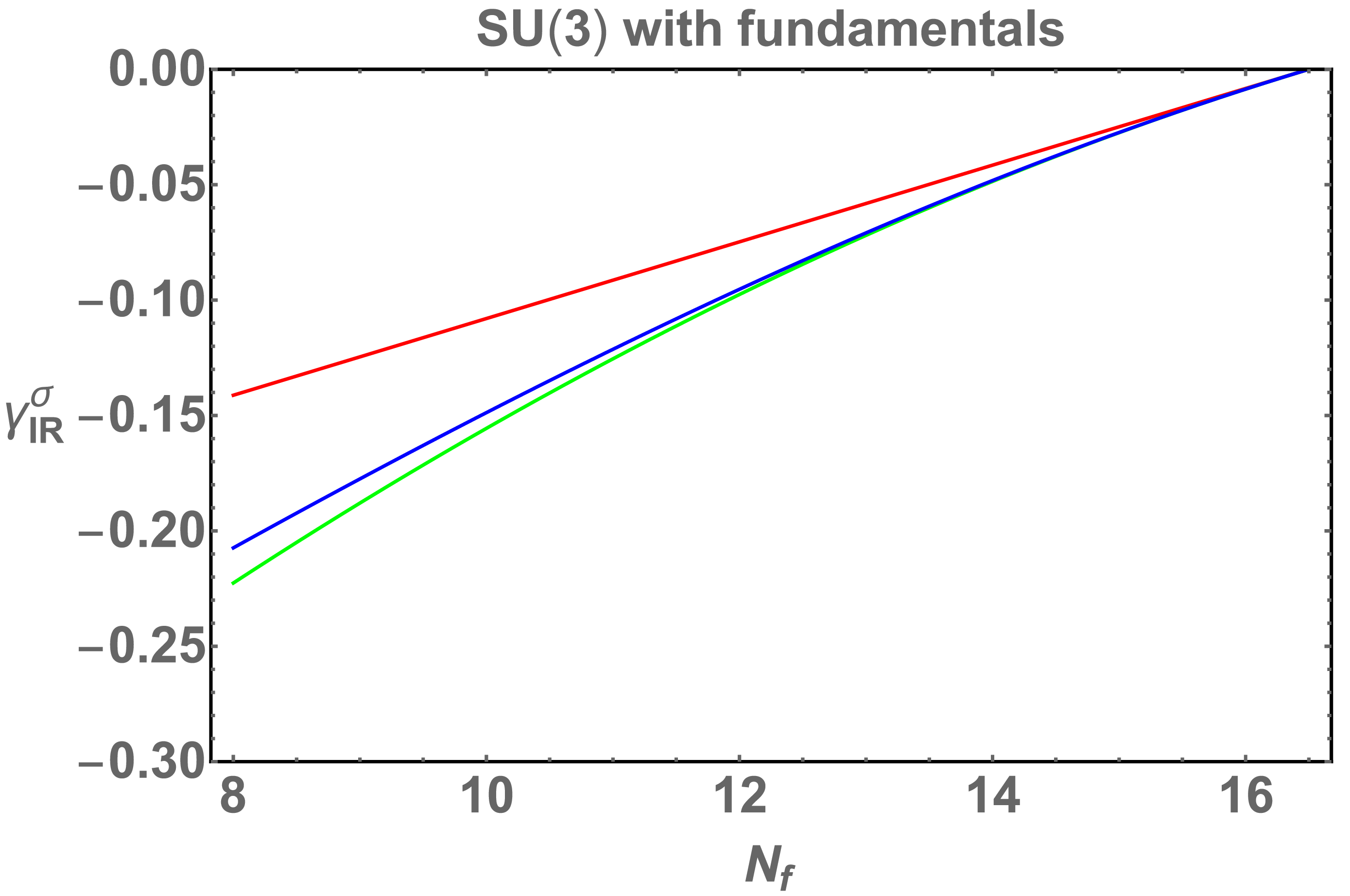}
  \end{center}
\caption{Plot of the anomalous dimension $\gamma^{(\sigma)}_{IR,F}$ of the 
operator $\bar\psi \sigma_{\lambda\mu_1} \psi$ at the IRFP for 
$G={\rm SU}(3)$, and $N_f$ fermions in the fundamental representation,
calculated to order $O(\Delta_f^p)$, where $p=1, \ 2, \ 3$. Denoting the 
calculation to order $O(\Delta_f^p)$ as 
$\gamma^{(\sigma)}_{IR,F,\Delta_f^p}$, from top to botton
(with colors online), the colors refer to 
$\gamma^{(\sigma)}_{IR,F,\Delta_f}$ (red), 
$\gamma^{(\sigma)}_{IR,F,\Delta_f^2}$ (green), and 
$\gamma^{(\sigma)}_{IR,F,\Delta_f^3}$ (blue).}
\label{s_fund_plot}
\end{figure}

\begin{figure}
  \begin{center}
    \includegraphics[height=6cm]{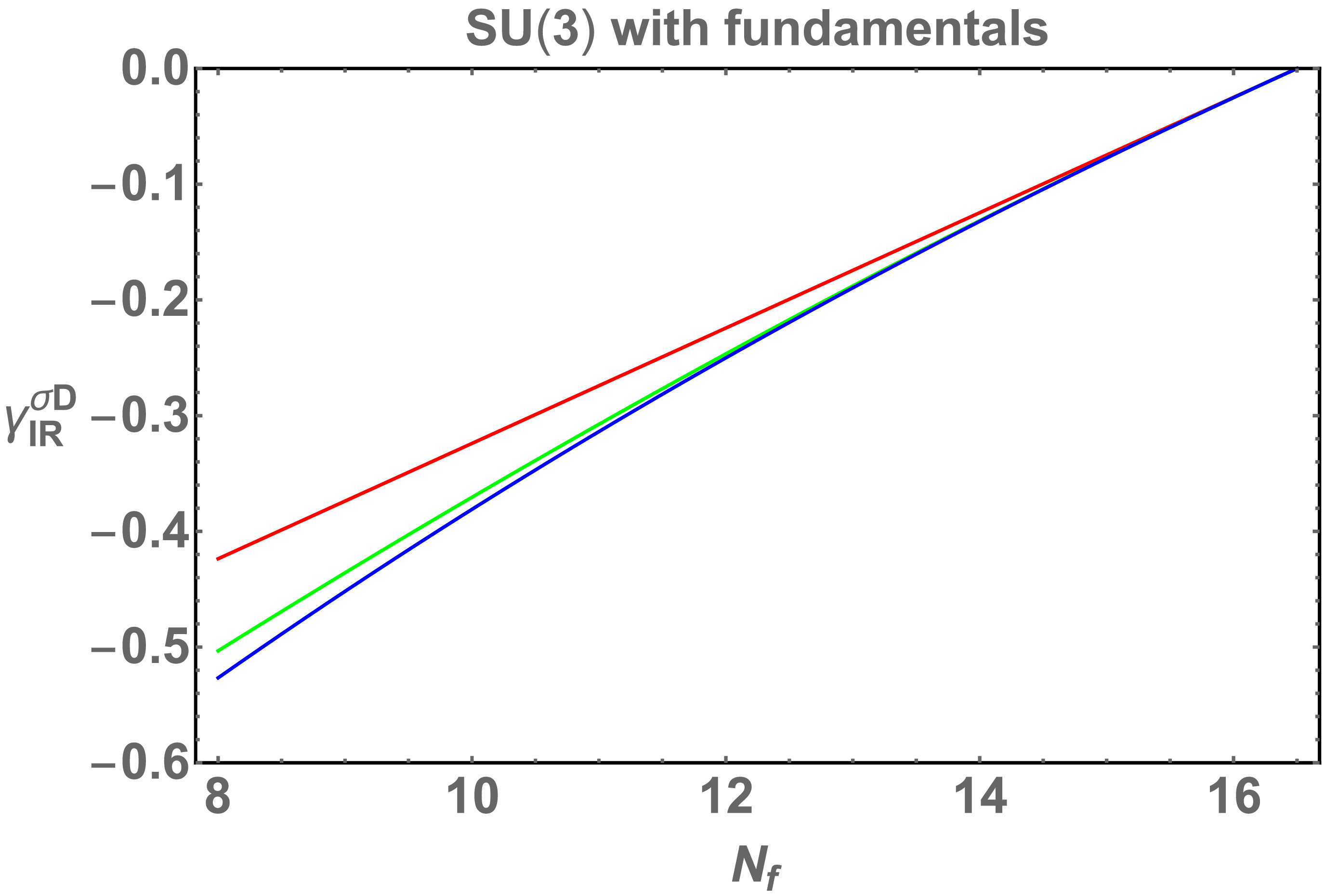}
  \end{center}
\caption{Plot of the anomalous dimension $\gamma^{(\sigma D)}_{IR,F}$ of the 
operator $\bar\psi \sigma_{\lambda\mu_1} D_{\mu_2} \psi$ at the IRFP for 
$G={\rm SU}(3)$, and $N_f$ fermions in the fundamental representation,
calculated to order $O(\Delta_f^p)$, where $p=1, \ 2, \ 3$. Denoting the 
calculation to order $O(\Delta_f^p)$ as 
$\gamma^{(\sigma D)}_{IR,F,\Delta_f^p}$, from top to botton
(with colors online), the colors refer to 
$\gamma^{(\sigma D)}_{IR,F,\Delta_f}$ (red), 
$\gamma^{(\sigma D)}_{IR,F,\Delta_f^2}$ (green), and 
$\gamma^{(\sigma D)}_{IR,F,\Delta_f^3}$ (blue).}
\label{sd_fund_plot}
\end{figure}

\begin{figure}
  \begin{center}
    \includegraphics[height=6cm]{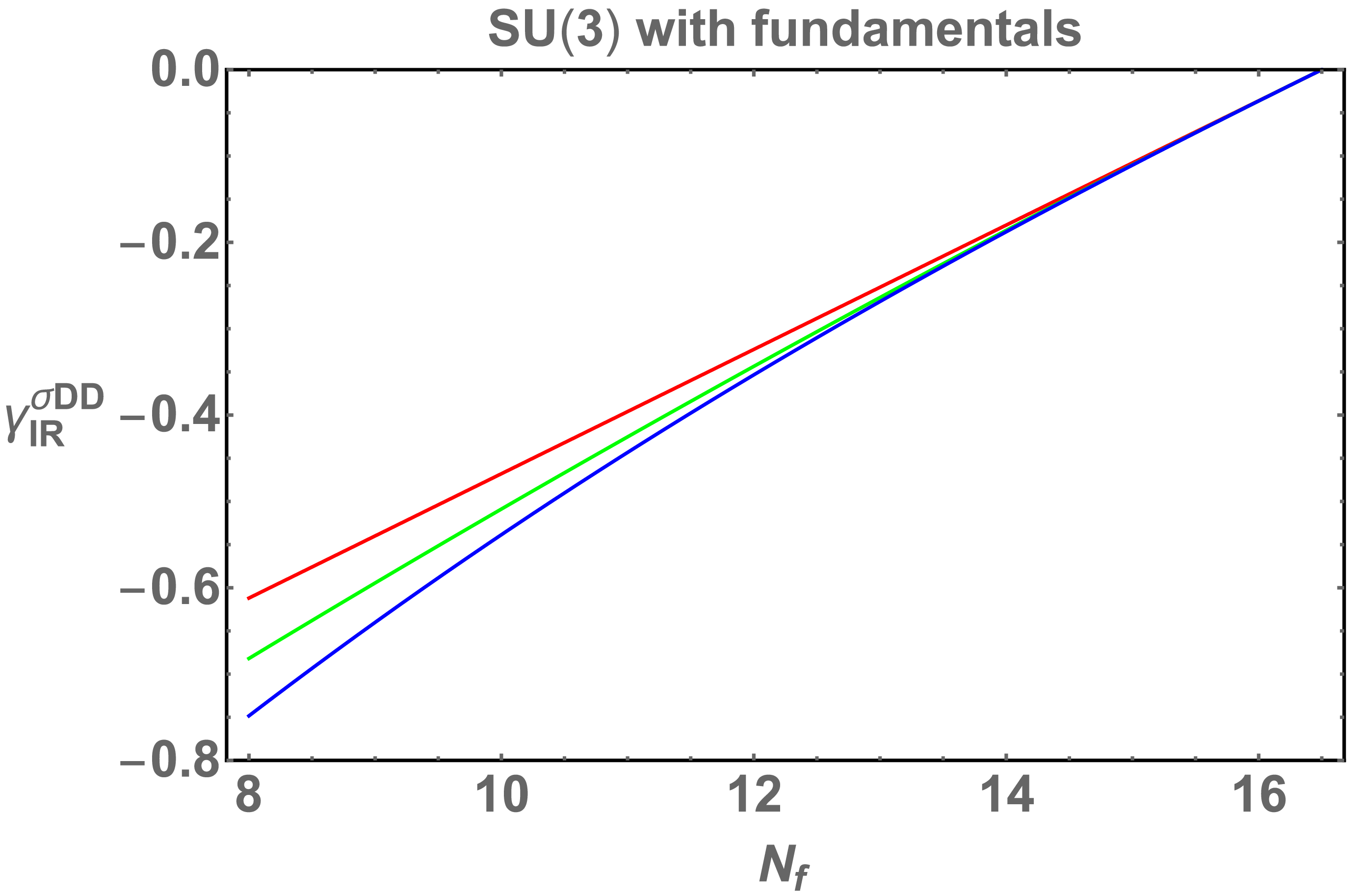}
  \end{center}
\caption{Plot of the anomalous dimension $\gamma^{(\sigma DD)}_{IR,F}$ of the 
operator $\bar\psi \sigma_{\lambda\mu_1} D_{\mu_2} D_{\mu_3}\psi$ at the 
IRFP for $G={\rm SU}(3)$, and $N_f$ fermions in the fundamental representation,
calculated to order $O(\Delta_f^p)$, where $p=1, \ 2, \ 3$. Denoting the 
calculation to order $O(\Delta_f^p)$ as 
$\gamma^{(\sigma DD)}_{IR,F,\Delta_f^p}$, from top to botton 
(with colors online), the colors refer to 
$\gamma^{(\sigma DD)}_{IR,F,\Delta_f}$ (red), 
$\gamma^{(\sigma DD)}_{IR,F,\Delta_f^2}$ (green), and 
$\gamma^{(\sigma DD)}_{IR,F,\Delta_f^3}$ (blue).} 
\label{sdd_fund_plot}
\end{figure}

\begin{figure}
  \begin{center}
    \includegraphics[height=6cm]{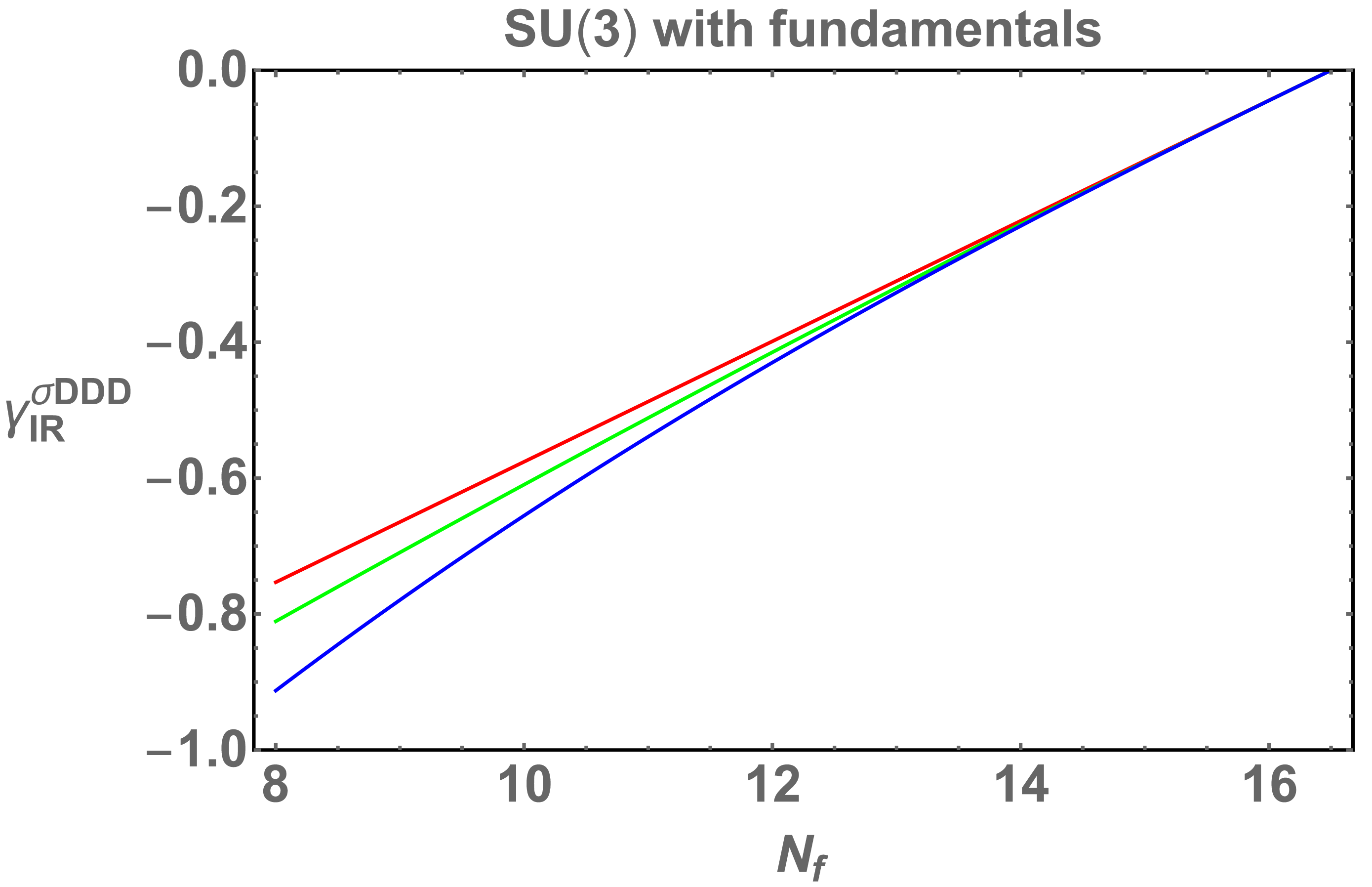}
  \end{center}
\caption{Plot of the anomalous dimension $\gamma^{(\sigma DDD)}_{IR,F}$ of the 
operator $\bar\psi \sigma_{\lambda\mu_1} D_{\mu_2} D_{\mu_2} D_{\mu_3} \psi$ 
at the IRFP for 
$G={\rm SU}(3)$, and $N_f$ fermions in the fundamental representation,
calculated to order $O(\Delta_f^p)$, where $p=1, \ 2, \ 3$. Denoting the 
calculation to order $O(\Delta_f^p)$ as 
$\gamma^{(\sigma DDD)}_{IR,F,\Delta_f^p}$, from top to bottom 
(with colors online), the colors refer to 
$\gamma^{(\sigma DDD)}_{IR,F,\Delta_f}$ (red), 
$\gamma^{(\sigma DDD)}_{IR,F,\Delta_f^2}$ (green), and 
$\gamma^{(\sigma DDD)}_{IR,F,\Delta_f^3}$ (blue).} 
\label{sddd_fund_plot}
\end{figure}

In Figs. \ref{gd_fund_plot}-\ref{sddd_fund_plot} we show plots of these
anomalous dimensions for the SU(3) theory with $R=F$. The plot of the anomalous
dimension for $\bar\psi\sigma_{\lambda\mu_1}\psi$ is based on the analytic
results of our earlier paper \cite{dex} but was not given there and is new
here.  As can be seen from these tables and figures, the higher-order terms in
the $\Delta_f$ expansion are sufficiently small that it is expected to be
reliable throughout much of the non-Abelian Coulomb phase (i.e., conformal
window).  As is obvious, since our calculations are finite series expansions in
powers of $\Delta_f$, they are most accurate in the upper part of the NACP,
where this expansion parameter $\Delta_f$ is small.  This is similar to what we
found in our earlier scheme-independent calculations of anomalous dimensions
\cite{gtr}-\cite{dexo}. In the figures, this is evident from the fact that the
curves for the anomalous dimensions calculated to $O(\Delta_f^3)$ are
reasonably close to the corresponding curves for these anomalous dimensions
calculated to order $O(\Delta_f^2)$.


\section{LNN Limit for $\gamma^{({\cal O})}_{IR,{\rm SU}(N_c),F}$ }
\label{gamma_op_lnn_section}

In a theory with gauge group SU($N_c$) and fermions in the fundamental
representation, $R=F$, it is of interest to consider the limit 
\beqs
& & N_c \to \infty \ , \quad N_F \to \infty  \quad {\rm with} \
r \equiv \frac{N_F}{N_c} \ {\rm fixed \ and \ finite}  \cr\cr
& & {\rm and} \ \ \xi(\mu) \equiv \alpha(\mu) N_c \ {\rm is \ a \
finite \ function \ of} \ \mu \ .
\cr\cr
& &
\label{lnn}
\eeqs
This limit is denoted as $\lim_{\rm LNN}$ (where ``LNN'' connotes 
``large $N_c$ and $N_F$'' with the constraints in Eq. (\ref{lnn}) imposed).
It is also often called the 't Hooft-Veneziano limit.  It has the
simplifying feature that rather than depending on $N_c$ and $N_f$, the 
properties of the theory only depend on their ratio, $r$. 
The scheme-independent expansion parameter in this LNN limit is 
\beq
\Delta_r \equiv \lim_{\rm LNN} \frac{\Delta_f}{N_c} = \frac{11}{2}-r \ .
\label{deltar}
\eeq
\beq
r_u = \lim_{\rm LNN} \frac{N_u}{N_c} \ ,
\label{rb1zdef}
\eeq
and
\beq
r_\ell = \lim_{\rm LNN} \frac{N_\ell}{N_c} \ ,
\label{rb2zdef}
\eeq
with values
\beq
r_u = \frac{11}{2} = 5.5
\label{rb1z}
\eeq
and
\beq
r_\ell = \frac{34}{13}=2.6154  \ .
\label{rb2z}
\eeq
With $I_{IRZ}: \ N_\ell < N_f < N_u$, it follows that the corresponding
interval in the ratio $r$ is
\beq
I_{IRZ,r}: \quad \frac{34}{13} < r < \frac{11}{2}, \ i.e.,
\ 2.6154 < r < 5.5
\label{intervalr}
\eeq

Here we evaluate these scheme-independent anomalous dimension coefficients 
in a theory with $G={\rm SU}(N_c)$ and $R=F$, in the LNN limit. 
The rescaled coefficients that are finite in the LNN limit are 
\beq
\hat \kappa^{({\cal O})}_n = \lim_{N_c \to \infty} N_c^n \kappa^{({\cal O})}_n
\label{kappa_op_hat}
\eeq
The anomalous dimension $\gamma^{({\cal O})}_{IR}$ is also finite in this 
limit and is given by
\beq
\lim_{\rm LNN} \gamma^{({\cal O})}_{IR,{\rm SU}(N_c),F} = 
\sum_{n=1}^\infty \kappa^{({\cal O})}_n \Delta_f^n 
= \sum_{n=1}^\infty \hat \kappa^{({\cal O})}_n \Delta_r^n \ . 
\label{gamma_ir_lnn}
\eeq
As $r$ decreases from its upper limit, $r_u$, to 
$r_\ell$, the expansion variable $\Delta_r$ increases from 0 to 
\beq
(\Delta_r)_{\rm max} = 
\frac{75}{26} = 2.8846 \quad {\rm for} \ r \in I_{IRZ,r} \ .
\label{deltarmax}
\eeq

In this LNN limit, the values of $\hat\kappa^{({\cal O})}_n$ with
$1 \le n \le 3$ for the operators ${\cal O}$ considered here are listed in
Table \ref{kappa_hat_values}. For comparison, we also include the corresponding
values of $\hat\kappa^{({\cal O})}_n$ for the operators $\bar\psi\psi$ and 
$\bar\psi\sigma_{\mu\nu}\psi$ that we had calculated in \cite{dex}. 


\section{Evaluation of Anomalous Dimensions $\gamma^{({\cal O})}_{IR}$
for $G={\rm SU}(N_c)$ and $R=Adj$}
\label{gamma_adj_section}

For the case where $G={\rm SU}(N_c)$ and the fermions are in the adjoint
representation, $R=Adj$, our general results for the scheme-independent 
expansion coefficients for the anomalous dimensions of the operators under
consideration are as follows:
:
\beq
\kappa^{(\gamma D)}_{1,{\rm SU}(N_c),Adj} = -\frac{2^5}{3^4} = -0.395062 \ , 
\label{kappa1_gamma_d_adj}
\eeq
\beq
\kappa^{(\gamma D)}_{2,{\rm SU}(N_c),Adj} = -\frac{1756}{3^9} = -0.0892140 \ , 
\label{kappa2_gamma_d_adj}
\eeq
\beqs
\kappa^{(\gamma D)}_{3,{\rm SU}(N_c),Adj} &=& 
-\frac{88129}{3^{14}} + \frac{4736}{3^{10}N_c^2} \cr\cr
& = & -0.0184256 + \frac{0.0802046}{N_c^2}  \ , 
\label{kappa3_gamma_d_adj}
\eeqs
%
\beq
\kappa^{(\gamma DD)}_{1,{\rm SU}(N_c),Adj} = -\frac{50}{3^4} = -0.617284
 \ , 
\label{kappa1_gamma_dd_adj}
\eeq
\beq
\kappa^{(\gamma DD)}_{2,{\rm SU}(N_c),Adj} = -\frac{10165}{2^2 \cdot 3^9} 
= -0.129109 \ , 
\label{kappa2_gamma_dd_adj}
\eeq
\beqs
\kappa^{(\gamma DD)}_{3,{\rm SU}(N_c),Adj} &=& 
-\frac{2272255}{2^4 \cdot 3^{14}} + \frac{7400}{3^{10} N_c^2 } \cr\cr
&=& -0.0296920 + \frac{0.125320}{N_c^2} \ ,
\label{kappa3_gamma_dd_adj}
\eeqs
%
\beq
\kappa^{(\gamma DDD)}_{1,{\rm SU}(N_c),Adj} = -\frac{314}{3^4 \cdot 5} =
-0.775309  \ , 
\label{kappa1_gamma_ddd_adj}
\eeq
\beq
\kappa^{(\gamma DDD)}_{2,{\rm SU}(N_c),Adj} = -\frac{1504769}{2^2 \cdot 3^9 \cdot
5^3} = -0.152900 \ , 
\label{kappa2_gamma_ddd_adj}
\eeq
and
\beqs
\kappa^{(\gamma DDD)}_{3,{\rm SU}(N_c),Adj} &=& 
-\frac{9206650603}{2^4 \cdot 3^{14} \cdot 5^5} + 
\frac{46472}{3^{10} \cdot 5 N_c^2 } \cr\cr
&=& -0.0384976 + \frac{0.1574015}{N_c^2} \ , 
\label{kappa3_gamma_ddd_adj}
\eeqs
%
\beq
\kappa^{(\sigma D)}_{1,{\rm SU}(N_c),Adj} = -\frac{2^2}{3^2} = -0.444444 \ , 
\label{kappa1_sigma_d_adj}
\eeq
\beq
\kappa^{(\sigma D)}_{2,{\rm SU}(N_c),Adj} = -\frac{149}{2 \cdot 3^6} = 
-0.102195 \ , 
\label{kappa2_sigma_d_adj}
\eeq
and
\beqs
\kappa^{(\sigma D)}_{3,{\rm SU}(N_c),Adj} &=& -\frac{10801}{2^3 \cdot 3^{10}} 
+ \frac{592}{3^8 N_c^2} \cr\cr
&=& -0.0228645 + \frac{0.0902301}{N_c^2} \ , 
\label{kappa3_sigma_d_adj}
\eeqs
%
\beq
\kappa^{(\sigma DD)}_{1,{\rm SU}(N_c),Adj} = -\frac{52}{3^4} = -0.641975 \ , 
\label{kappa1_sigma_dd_adj}
\eeq
\beq
\kappa^{(\sigma DD)}_{2,{\rm SU}(N_c),Adj} = -\frac{5123}{2 \cdot 3^9} = 
-0.130138  \ , 
\label{kappa2_sigma_dd_adj}
\eeq
\beqs
\kappa^{(\sigma DD)}_{3,{\rm SU}(N_c),Adj} &=& 
-\frac{984949}{2^3 \cdot 3^{14}} + \frac{7696}{3^{10}N_c^2 } \cr\cr
&=& -0.0257410 + \frac{0.130332}{N_c^2} \ ,
\label{kappa3_sigma_dd_adj}
\eeqs
%
\beq
\kappa^{(\sigma DDD)}_{1,{\rm SU}(N_c),Adj} = -\frac{2^6}{3^4} = -0.790123  
\ , 
\label{kappa1_sigma_ddd_adj}
\eeq
\beq
\kappa^{(\sigma DDD)}_{2,{\rm SU}(N_c),Adj} = -\frac{3070}{3^9} = -0.155972
\ , 
\label{kappa2_sigma_ddd_adj}
\eeq
and
\beqs
\kappa^{(\sigma DDD)}_{3,{\rm SU}(N_c),Adj} &=& 
-\frac{378247}{2 \cdot 3^{14}} + \frac{9472}{3^{10} N_c^2 } \cr\cr
&=& -0.0395410 + \frac{0.160409}{N_c^2} \ . 
\label{kappa3_sigma_ddd_adj}
\eeqs

For all of these operators ${\cal O}$, the coefficients $\kappa^{({\cal
    O})}_{n,{\rm SU}(N_c),Adj}$ are negative for $n=1$ and $n=2$ and for all
$N_c$.  The coefficient $\kappa^{(\sigma D)}_{3,{\rm SU}(N_c),Adj}$ is
  negative for all $N_c$, while the coefficients 
$\kappa^{({\cal O})}_{3,{\rm SU}(N_c),Adj}$ for
  the other operators are positive for $N_c=2$, i.e., $G={\rm SU}(2)$, and are
  negative for $N_c \ge 3$.


%
\section{Conclusions}
\label{conclusion_section}

In conclusion, in this paper we have calculated scheme-independent expansions
up to $O(\Delta_f^3)$ inclusive for the anomalous dimensions of the
higher-spin, twist-2 bilinear fermion 
operators $\bar \psi \gamma_{\mu_1} D_{\mu_2}...D_{\mu_j}
\psi$ and $\bar \psi \sigma_{\lambda\mu_1} D_{\mu_2}...D_{\mu_j} \psi$ with $j$
up to 3, evaluated at an IR fixed point in the non-Abelian Coulomb phase of an
asymptotically free gauge theory with gauge group $G$ and $N_f$ fermions
transforming according to a representation $R$ of $G$.  Our general results are
evaluated for several special cases, including the case $G={\rm SU}(N_c)$ with
$R$ equal to the fundamental and adjoint representations.  We have presented
our results in convenient tabular and graphical formats.  For fermions in the
fundamental representation, we also analyze the limit $N_c \to \infty$ and $N_f
\to \infty$ with $N_f/N_c$ fixed and finite.  A comparison with our previous
scheme-independent calculations of the corresponding anomalous dimensions of
$\bar\psi\psi$ and $\bar\psi \sigma_{\mu\nu}\psi$ has also been given. Our new
results further elucidate the properties of conformal field theories.  With the
requisite inputs, one could extend these scheme-independent calculations to
higher-spin operators and to higher order in powers of $\Delta_f$. It is hoped
that lattice measurements of these anomalous dimensions of higher-spin
operators in the conformal window will be performed in the future, and it will
be of interest to compare our calculations with lattice results when they will
become available.


\begin{acknowledgments}

This research was supported in part by the Danish National
Research Foundation grant DNRF90 to CP$^3$-Origins at SDU (T.A.R.) and
by the U.S. NSF Grants NSF-PHY-1620628 and NSF-PHY-1915093 (R.S.). 

\end{acknowledgments}


\bigskip
\bigskip

\begin{appendix}

\section{Previous Results on $\gamma^{(1)}$ and 
$\gamma^{(\sigma)}$ for $G={\rm SU}(3)$ and $R=F$ }
\label{gammat_su3_appendix}

In this appendix, for comparison with our new results, we list our previous
results from \cite{dex} (see also \cite{dexl}) 
for the scheme-independent series expansions of the anomalous
dimensions $\gamma^{(\cal O)}_{IR}$ for 
${\cal O} = \bar\psi\psi$ and 
${\cal O} = \bar\psi \sigma_{\mu\nu}\psi$.  Following the same shorthand
notation as in the text, we denote the coefficients 
at order $O(\Delta_f^n)$ in the scheme-independent series expansions 
(\ref{gamma_delta_series}) for these anomalous dimensions as 
$\kappa^{(1)}_n$ and $\kappa^{(\sigma)}_n$.  We calculated 
\beq
\kappa^{(1)}_1 = \frac{8T_fC_f}{C_AD} \ ,
\label{kappa1}
\eeq
\beq
\kappa^{(1)}_2 = \frac{4T_f^2C_f(5C_A+88C_f)(7C_A+4C_f)}{3C_A^2 D^3} \ ,
\label{kappa2}
\eeq
and
\begin{widetext}
\beqs
\kappa^{(1)}_3 &=& \frac{4T_fC_f}{3^4 C_A^4 D^5} \bigg [
-55419T_f^2C_A^5 + 432012T_f^2C_A^4C_f
+ 5632T_f^2 C_f \, \frac{d_A^{abcd}d_A^{abcd}}{d_A} \, (-5+132\zeta_3) \cr\cr
&+&
16C_A^3 \bigg ( 122043T_f^2 C_f^2 + 6776 \, \frac{d_R^{abcd}d_R^{abcd}}{d_A}
\, (-11+24\zeta_3) \bigg ) \cr\cr
&+& 704C_A^2 \bigg ( 1521 T_f^2 C_f^3 + 112 T_f \,
\frac{d_R^{abcd}d_A^{abcd}}{d_A} \, (4-39\zeta_3)
+ 242C_f \, \frac{d_R^{abcd}d_R^{abcd}}{d_A} \, (-11+24\zeta_3) \bigg )
\cr\cr
&+& 32T_fC_A \bigg ( 53361T_fC_f^4 - 3872 C_f
\, \frac{d_R^{abcd}d_A^{abcd}}{d_A} \, (-4+39\zeta_3)
+ 112T_f \, \frac{d_A^{abcd}d_A^{abcd}}{d_A} \, (-5+132\zeta_3) \bigg )
\bigg ] 
\label{kappa3}
\eeqs
\end{widetext}
(where the denominator factor $D$ was defined in Eq. (\ref{d})). 
In \cite{dexl,dexo} we presented results for the next-higher order
coefficient, $\kappa^{(1)}_4$, but these are not needed here. 

For the $\kappa^{(\sigma)}_n$ we found 
\beq
\kappa^{(\sigma)}_1 = -\frac{8C_fT_f}{3C_A D}
\label{kappa_s_1}
\eeq
\beq
\kappa^{(\sigma)}_2 = 
-\frac{4C_fT_f^2(259C_A^2+428C_AC_f-528C_f^2)}{9C_A^2 D^3} 
\label{kappa_s_2}
\eeq
and
\begin{widetext}
\beqs
\kappa^{(\sigma)}_3 
& = & \frac{4C_f T_f}{3^5 C_A^4 D^5}\bigg [ 3C_A T_f^2 \bigg \{ 
     C_A^4(-11319+188160\zeta_3) + 
  C_A^3C_f(-337204+64512\zeta_3) + C_A^2C_f^2(83616-890112\zeta_3) \cr\cr
&+& C_AC_f^3(1385472-354816\zeta_3) + C_f^4(-212960+743424\zeta_3) \bigg \}
-512T_f^2D(-5+132\zeta_3)\frac{d_A^{abcd}d_A^{abcd}}{d_A} \cr\cr
&-& 15488C_A^2D(-11+24\zeta_3)\frac{d_R^{abcd}d_R^{abcd}}{d_A}
+11264C_AT_f D(-4+39\zeta_3)\frac{d_R^{abcd}d_A^{abcd}}{d_A} \bigg ]  \ . 
\label{kappa_s_3}
\eeqs
\end{widetext}

For $G={\rm SU}(N_c)$ and $R=F$, in the LNN limit, these yield the rescaled
coefficients 
\beq
\hat\kappa^{(1)}_1 = \frac{4}{5^2} = 0.1600 \ ,
\label{kappahat_1_1}
\eeq
\beq
\hat\kappa^{(1)}_2 = \frac{588}{5^6} = 0.037632 \ ,
\label{kappahat_1_2}
\eeq
\beq
\hat\kappa^{(1)}_3 = 
\frac{2193944}{3^3 \cdot 5^{10}} = 0.83207 \times 10^{-2} \ ,
\label{kappahat_1_3}
\eeq
\beq
\hat \kappa^{(\sigma)}_1 = -\frac{4}{3 \cdot 5^2} = -0.053333 \ ,
\label{kappahat_s_1}
\eeq
\beq
\hat \kappa^{(\sigma)}_2 = -\frac{1364}{3^2 \cdot 5^6} = 
-(0.969956 \times 10^{-2}) \ ,
\label{kappahat_s_2}
\eeq
and
\beq
\hat \kappa^{(\sigma)}_3 = \frac{184456}{3^4 \cdot 5^{10}} = 
2.3319 \times 10^{-4} \ .
\label{kappahat_s_3}
\eeq

\end{appendix} 



\newpage

\begin{table}
\caption{\footnotesize{Signs of scheme-independent 
expansion coefficients $\kappa^{({\cal O})}_n$ for 
gauge group $G={\rm SU}(N_c)$ with $N_ c \ge 2$ and fermion 
representation $R=F$ (fundamental).}}
\begin{center}
\begin{tabular}{|c|c|c|c|} \hline\hline
${\cal O}$ & 
$\kappa^{({\cal O})}_{1,{\rm SU}(N_c),F}$ & 
$\kappa^{({\cal O})}_{2,{\rm SU}(N_c),F}$ & 
$\kappa^{({\cal O})}_{3,{\rm SU}(N_c),F}$ \\ \hline
$\bar\psi\psi$                                      & $+$  & $+$  & $+$ \\

$\bar\psi \sigma_{\lambda\mu_1}\psi$                & $-$  & $-$  & $+$ \\

$\bar\psi \gamma_{\mu_1} D_{\mu_2} \psi$            & $-$  & $-$  & $-$ \\

$\bar\psi \gamma_{\mu_1} D_{\mu_2} D_{\mu_3} \psi$  & $-$  & $-$  & $-$ \\

$\bar\psi \gamma_{\mu_1} D_{\mu_2}D_{\mu_3}D_{\mu_4}\psi$&$-$ & $-$ & $-$ \\

$\bar\psi \sigma_{\lambda\mu_1}D_{\mu_2}\psi$       & $-$   & $-$  & $-$   \\

$\bar\psi \sigma_{\lambda\mu_1}D_{\mu_2}D_{\mu_3}\psi$ & $-$ & $-$  & $-$ \\

$\bar\psi \sigma_{\lambda\mu_1}D_{\mu_2}D_{\mu_3}D_{\mu_4}\psi$
                                                       & $-$ & $-$ & $-$ \\
\hline\hline
\end{tabular}
\end{center}
\label{kappa_signs}
\end{table}


\begin{table}
  \caption{\footnotesize{
Values of the anomalous dimension 
$\gamma^{(\gamma D)}_{IR,F}$ calculated to $O(\Delta_f^p)$, denoted 
$\gamma^{(\gamma D)}_{IR,F,\Delta_f^p}$, with
$1 \le p \le 3$, for $G={\rm SU}(3)$, as a function of $N_f$.}}
\begin{center}
\begin{tabular}{|c|c|c|c|} \hline\hline
$N_f$ & 
$\gamma^{(\gamma D)}_{IR,F,\Delta_f}$    &
$\gamma^{(\gamma D)}_{IR,F,\Delta_f^2}$  &
$\gamma^{(\gamma D)}_{IR,F,\Delta_f^3}$  
\\ \hline
8  & $-0.377$   & $-0.446$  & $-0.481$    \\
9  & $-0.332$   & $-0.386$  & $-0.411$    \\
10 & $-0.288$   & $-0.328$  & $-0.344$    \\
11 & $-0.244$   & $-0.273$  & $-0.282$    \\ 
12 & $-0.199$   & $-0.219$  & $-0.224$    \\ 
13 & $-0.155$   & $-0.167$  & $-0.169$    \\
14 & $-0.111$   & $-0.117$  & $-0.118$    \\
15 & $-0.0665$  & $-0.0686$ & $-0.0688$   \\
16 & $-0.02215$ & $-0.0224$ & $-0.0224$   \\
\hline\hline
\end{tabular}
\end{center}
\label{gamma_gd_values}
\end{table}


\begin{table}
  \caption{\footnotesize{
Values of the anomalous dimension 
$\gamma^{(\gamma DD)}_{IR,F}$ calculated to $O(\Delta_f^p)$, denoted 
$\gamma^{(\gamma DD)}_{IR,F,\Delta_f^p}$, with
$1 \le p \le 3$, for $G={\rm SU}(3)$, as a function of $N_f$.}}
\begin{center}
\begin{tabular}{|c|c|c|c|} \hline\hline
$N_f$ & 
$\gamma^{(\gamma DD)}_{IR,F,\Delta_f}$  &
$\gamma^{(\gamma DD)}_{IR,F,\Delta_f^2}$ &
$\gamma^{(\gamma DD)}_{IR,F,\Delta_f^3}$ 
\\ \hline
8  & $-0.588$  & $-0.654$  & $-0.724$     \\
9  & $-0.519$  & $-0.570$  & $-0.618$  \\
10 & $-0.450$  & $-0.488$  & $-0.520$  \\
11 & $-0.381$  & $-0.408$  & $-0.427$   \\ 
12 & $-0.3115$ & $-0.330$  & $-0.340$   \\ 
13 & $-0.242$  & $-0.253$  & $-0.258$   \\
14 & $-0.173$  & $-0.179$  & $-0.180$   \\
15 & $-0.104$  & $-0.106$  & $-0.106$   \\
16 & $-0.0346$ & $-0.0348$ & $-0.0349$  \\
\hline\hline
\end{tabular}
\end{center}
\label{gamma_gdd_values}
\end{table}


\begin{table}
  \caption{\footnotesize{
Values of the anomalous dimension 
$\gamma^{(\gamma DDD)}_{IR,F}$ calculated to $O(\Delta_f^p)$, denoted 
$\gamma^{(\gamma DDD)}_{IR,F, \Delta_f^p}$, with
$1 \le p \le 3$, for $G={\rm SU}(3)$, as a function of $N_f$.}}
\begin{center}
\begin{tabular}{|c|c|c|c|} \hline\hline
$N_f$ & 
$\gamma^{(\gamma DDD)}_{IR,F,\Delta_f}$  &
$\gamma^{(\gamma DDD)}_{IR,F,\Delta_f^2}$ &
$\gamma^{(\gamma DDD)}_{IR,F,\Delta_f^3}$ 
\\ \hline
8  & $-0.739$  & $-0.794$  & $-0.900$  \\
9  & $-0.652$  & $-0.695$  & $-0.7675$ \\
10 & $-0.565$  & $-0.598$  & $-0.645$  \\
11 & $-0.478$  & $-0.501$  & $-0.530$  \\ 
12 & $-0.391$  & $-0.407$  & $-0.422$  \\ 
13 & $-0.304$  & $-0.314$  & $-0.321$  \\
14 & $-0.217$  & $-0.222$  & $-0.225$  \\
15 & $-0.130$  & $-0.132$  & $-0.133$  \\
16 & $-0.0435$ & $-0.0437$ & $-0.0437$ \\
\hline\hline
\end{tabular}
\end{center}
\label{gamma_gddd_values}
\end{table}


\begin{table}
  \caption{\footnotesize{
Values of the anomalous dimension 
$\gamma^{(\sigma)}_{IR,F}$ 
calculated to $O(\Delta_f^p)$, denoted 
$\gamma^{(\sigma)}_{IR, F, \Delta_f^p}$, with
$1 \le p \le 3$, for $G={\rm SU}(3)$, as a function of $N_f$.}}
\begin{center}
\begin{tabular}{|c|c|c|c|} \hline\hline
$N_f$ & 
$\gamma^{(\sigma)}_{IR,F,\Delta_f}$  &
$\gamma^{(\sigma)}_{IR,F,\Delta_f^2}$ &
$\gamma^{(\sigma)}_{IR,F,\Delta_f^3}$ 
\\ \hline
8  & $-0.141$  & $-0.223$  & $-0.207$  \\
9  & $-0.125$  & $-0.188$  & $-0.1775$ \\
10 & $-0.108$  & $-0.156$  & $-0.149$  \\
11 & $-0.0914$ & $-0.125$  & $-0.121$  \\ 
12 & $-0.0748$ & $-0.0976$ & $-0.0953$ \\ 
13 & $-0.05815$& $-0.07195$& $-0.0709$  \\
14 & $-0.0415$ & $-0.0486$ & $-0.0482$  \\
15 & $-0.0249$ & $-0.0275$ & $-0.0274$  \\
16 & $-0.00831$& $-0.00859$& $-0.00859$ \\
\hline\hline
\end{tabular}
\end{center}
\label{gamma_s_values}
\end{table}


\begin{table}
  \caption{\footnotesize{
Values of the anomalous dimension 
$\gamma^{(\sigma D)}_{IR,F}$ calculated to $O(\Delta_f^p)$, denoted 
$\gamma^{(\sigma D)}_{IR, F, \Delta_f^p}$, with
$1 \le p \le 3$, for $G={\rm SU}(3)$, as a function of $N_f$.}}
\begin{center}
\begin{tabular}{|c|c|c|c|} \hline\hline
$N_f$ & 
$\gamma^{(\sigma D)}_{IR,F,\Delta_f}$  &
$\gamma^{(\sigma D)}_{IR,F,\Delta_f^2}$ &
$\gamma^{(\sigma D)}_{IR,F,\Delta_f^3}$ 
\\ \hline
8  & $-0.424$  & $-0.503$  & $-0.527$  \\
9  & $-0.374$  & $-0.436$  & $-0.452$ \\
10 & $-0.324$  & $-0.3705$ & $-0.381$  \\
11 & $-0.274$  & $-0.307$  & $-0.314$  \\ 
12 & $-0.224$  & $-0.247$  & $-0.250$  \\ 
13 & $-0.174$  & $-0.188$  & $-0.190$  \\
14 & $-0.125$  & $-0.131$  & $-0.132$  \\
15 & $-0.0748$ & $-0.0772$ & $-0.0774$  \\
16 & $-0.0249$ & $-0.0252$ & $-0.0252$ \\
\hline\hline
\end{tabular}
\end{center}
\label{gamma_sd_values}
\end{table}


\begin{table}
  \caption{\footnotesize{
Values of the anomalous dimension 
$\gamma^{(\sigma DD)}_{IR,F}$ calculated to $O(\Delta_f^p)$, denoted 
$\gamma^{(\sigma DD)}_{IR,F, \Delta_f^p}$, with
$1 \le p \le 3$, for $G={\rm SU}(3)$, as a function of $N_f$.}}
\begin{center}
\begin{tabular}{|c|c|c|c|} \hline\hline
$N_f$ & 
$\gamma^{(\sigma DD)}_{IR,F,\Delta_f}$  &
$\gamma^{(\sigma DD)}_{IR,F,\Delta_f^2}$ &
$\gamma^{(\sigma DD)}_{IR,F,\Delta_f^3}$ 
\\ \hline
8  & $-0.612$  & $-0.682$  & $-0.748$  \\
9  & $-0.540$  & $-0.594$  & $-0.640$ \\
10 & $-0.468$  & $-0.509$  & $-0.539$  \\
11 & $-0.396$  & $-0.425$  & $-0.443$  \\ 
12 & $-0.324$  & $-0.344$  & $-0.353$  \\ 
13 & $-0.252$  & $-0.264$  & $-0.268$  \\
14 & $-0.180$  & $-0.186$  & $-0.188$  \\
15 & $-0.108$  & $-0.110$  & $-0.111$  \\
16 & $-0.0360$ & $-0.03624$ & $-0.03625$ \\
\hline\hline
\end{tabular}
\end{center}
\label{gamma_sdd_values}
\end{table}


\begin{table}
  \caption{\footnotesize{
Values of the anomalous dimension 
$\gamma^{(\sigma DDD)}_{IR,F}$ calculated to $O(\Delta_f^p)$, denoted 
$\gamma^{(\sigma DDD)}_{IR,F, \Delta_f^p}$, with
$1 \le p \le 3$, for $G={\rm SU}(3)$, as a function of $N_f$.}}
\begin{center}
\begin{tabular}{|c|c|c|c|} \hline\hline
$N_f$ & 
$\gamma^{(\sigma DDD)}_{IR,F,\Delta_f}$  &
$\gamma^{(\sigma DDD)}_{IR,F,\Delta_f^2}$ &
$\gamma^{(\sigma DDD)}_{IR,F,\Delta_f^3}$ 
\\ \hline
8  & $-0.753$  & $-0.811$  & $-0.913$  \\
9  & $-0.665$  & $-0.709$  & $-0.779$ \\
10 & $-0.576$  & $-0.610$  & $-0.655$  \\
11 & $-0.487$  & $-0.511$  & $-0.539$  \\ 
12 & $-0.399$  & $-0.415$  & $-0.430$  \\ 
13 & $-0.310$  & $-0.320$  & $-0.327$  \\
14 & $-0.222$  & $-0.2265$ & $-0.229$  \\
15 & $-0.133$  & $-0.135$  & $-0.135$  \\
16 & $-0.0443$ & $-0.0445$ & $-0.0445$ \\
\hline\hline
\end{tabular}
\end{center}
\label{gamma_sddd_values}
\end{table}


\begin{table}
  \caption{\footnotesize{Values of the $\hat\kappa^{({\cal O})}_n$ 
coefficients for $G={\rm SU}(N_c)$ and $R=F$ in the LNN limit. The
operators are indicated by their shorthand symbols, so $1$ refers to 
$\bar\psi\psi$; $\sigma$ refers to $\bar\psi \sigma_{\lambda\mu_1}\psi$;
$\gamma D$ to $\bar\psi \gamma_{\mu_1} D_{\mu_2} \psi$, etc. The
notation $a$e-$n$ means $a \times 10^{-n}$.}}
\begin{center}
\begin{tabular}{|c|c|c|c|} \hline\hline
${\cal O}$ & 
$\hat\kappa^{({\cal O})}_1$  &
$\hat\kappa^{({\cal O})}_2$  &
$\hat\kappa^{({\cal O})}_3$  
\\ \hline
1             & 0.160000       & 0.0376320        &  0.832074e-2      \\

$\sigma$      & $-0.0533333$   & $-0.969956$e-2   & 2.33189e-4  \\
\hline

$\gamma D$    & $-0.142222$    & $-1.04844$e-2    &  $-2.135375$e-3 \\

$\gamma DD$   &  $-0.222222$    & $-1.11967$e-2    & $-0.404507$e-2  \\

$\gamma DDD$  & $-0.279111$    & $-1.08149$e-2    & $-0.572019$e-2    \\
\hline

$\sigma D$    & $-0.160000$    & $-0.0120320$     & $-1.967385$e-3   \\

$\sigma DD$   & $-0.231111$    & $-1.17960$e-2     & $-0.397447$e-2   \\

$\sigma DDD$  & $-0.284444$   & $-1.114495$e-2     & $-0.567795$e-2    \\
\hline\hline
\end{tabular}
\end{center}
\label{kappa_hat_values}
\end{table}


%


%


%

 
%


%


%



\begin{thebibliography}{99}
\bibitem{fm}
%
  This assumption of massless fermions does not entail any loss of generality,
  since a fermion with nonzero mass $m$ would be integrated out of the
  low-energy effective field theory that describes the physics at Euclidean
  momentum scales $\mu < m$ and hence would not affect the infrared limit $\mu
  \to 0$ that we consider here.
\bibitem{ope_gen}
%
There is an extensive literature on operator product expansions, the
renormalization group, and applications to deep inelastic scattering and QCD. 
Some of the early papers, in addition to \cite{wilson_ope}, include
\cite{cg69}-\cite{b1p}; early reviews include 
\cite{politzer74}-\cite{gross75}.
\bibitem{ff}
There are also twist-2 operators that are bilinear in the field-strength 
tensors of the gauge field and do not directly involve the
fermions. Here we focus on higher-spin bilinear fermion operators.
\bibitem{nfintegral}
%
Here and below, we will often formally generalize $N_f$ from a positive
integer to a positive real number.  If an expression such as (\ref{Nfb1z}) 
or (\ref{Nfb2z}) evaluates to a non-integral real value, it is understood
implicitly that one infers an appropriate integral value from it.
\bibitem{group_invariants}
%
  We recall the definitions of relevant group invariants.  Let $T_a$ denote the
  generators of a Lie algebra of $G$, and let $d(R)$ denote the dimension of a
  representation $R$ of $G$. Then the quadratic Casimir invariant $C_2(R)$ is
  defined by $\sum_a T^a_RT^a_R = C_2(R) I$, where $I$ is the $d_R \times d_R$
  identity matrix. We utilize the common notation $C_2(Adj) \equiv C_2(G) 
  \equiv C_A$. The trace invariant $T(R)$ is defined by ${\rm Tr}_R(T^a_R
  T^b_R) = T(R)\delta^{ab}$. We use standard normalizations for SU($N_c$) so
  that, e.g., for the fundamental representation $F$, $T(F)=1/2$ and
  $C_2(F)=(N_c^2-1)/(2N_c)$.
\bibitem{gammaconvention}
%
Some authors use the opposite sign convention for the anomalous dimension,
writing $D_{\cal O} = D_{\cal O,{\rm free}} + \gamma_{\cal O}$. Our sign
convention is the same as the one used in the lattice gauge theory 
literature \cite{lgtreviews,simons}. The equivalent notation 
$\gamma^{({\cal O})} \equiv \gamma_{\cal O}$ will often be used.
\bibitem{sksb}
%
In contrast, in the confined QCD-type phase at lower $N_f$,
although the UV beta function may exhibit a formal IR zero, as the theory flows
from the UV to the IR, the running coupling becomes large enough so that 
spontaneous chiral symmetry breaking occurs, giving the fermions dynamical
masses, so that they are integrated out of the low-energy effective field
theory that is applicable deeper in the IR, and hence the resultant beta
function is that of a pure gauge theory, which does not have a (perturbative)
IR zero.  If $N_f$ is only slightly less than $N_{f,cr}$, these theories
thus exhibit quasi-conformal behavior. If the boundary in $N_f$ between 
the NACP and the QCD-type regime involves a continuous, although non-analytic,
change in the values of anomalous dimensions, then our previous
scheme-independent results in \cite{gtr}-\cite{dexo} and our results here may
give some approximate guide to the values of the corresponding 
anomalous dimensions in quasi-conformal theories with $N_f$ slightly less than
$N_{f,cr}$.
\bibitem{lowerend}
%
  Some of the lowest values of $N_f$ in these tables and figures may lie below
  $N_{f,cr}$.  However, since at present there is not a complete consensus
  among lattice groups concerning the value of $N_{f,cr}$ for this SU(3) theory
  with fermions in the fundamental representation, $R=F$
  \cite{lgtreviews,simons}, we list these values for completeness.
\end{thebibliography}
\end{document}